\documentclass[pra,twocolumn,notitlepage,superscriptaddress,nofootinbib,longbibliography]{revtex4-1}
\usepackage[english]{babel}
\usepackage{amsmath,amssymb,bbm,mathrsfs,color,graphicx,bm,braket}
\usepackage[colorlinks,citecolor=blue,urlcolor=blue]{hyperref}

\newcommand{\abs}[1]{\left\lvert #1 \right\rvert}

\newcommand{\unitvec}[1]{\hat{\mathbf{#1}}}

\def \be{\begin{equation}}
\def \ee{\end{equation}}
\def \ba{\begin{array}}
\def \ea{\end{array}}
\def \bea{\begin{eqnarray}}
\def \eea{\end{eqnarray}}

\def \half{{1\over 2}}

\def \t{{\theta}}

\def \D{{\Delta}}
\def \d{{\delta}}

\def \av#1{{\langle#1\rangle}}
\def\bnabla{\bm{\nabla}}
\def\bv{{\bf v}}
\def\bzh{{\bf \hat{z}}}
\def\br{{\bf r}}
\def\bp{{\bf p}}
\def\bP{{\bf P}}

\def\bB{{\bf B}}
\def\bE{{\bf E}}
\def\bJ{{\bf J}}
\def\bA{{\bf A}}
\def\bxi{\bm{\xi}}
\def\D{\mathcal{D}}

\def \be{\begin{equation}}
\def \ee{\end{equation}}
\def \bw{\begin{widetext}}
\def \ew{\end{widetext}}
\def \ba{\begin{array}}
\def \ea{\end{array}}
\def \bea{\begin{eqnarray}}
\def \eea{\end{eqnarray}}

\def \half{\frac{1}{2}}

\def \br{{\bf r}}

\def \t{{\theta}}

\def \D{{\Delta}}
\def \d{{\delta}}

\def \av#1{{\langle#1\rangle}}

\begin{document}

\title{Electrodynamic duality and vortex unbinding in driven-dissipative condensates}

\author{G. Wachtel} \affiliation{Department of Condensed Matter Physics,
  Weizmann Institute of Science, Rehovot 7610001, Israel}
\affiliation{Department of Physics, University of Toronto, Toronto, Ontario M5S
  1A7, Canada}

\author{L. M. Sieberer}
\affiliation{Department of Condensed Matter Physics, Weizmann Institute of Science,
  Rehovot 7610001, Israel}

\affiliation{Department of Physics, University of California, Berkeley,
  California 94720, USA}

\author{S. Diehl}
\affiliation{Institute of Theoretical Physics, University of Cologne, D-50937
  Cologne, Germany}

\author{E. Altman}
\affiliation{Department of Condensed Matter Physics, Weizmann Institute of Science,
  Rehovot 7610001, Israel}

\affiliation{Department of Physics, University of California, Berkeley,
  California 94720, USA}

% \author{Gideon Wachtel$^1$, Lukas Sieberer$^2$, Sebastian Diehl$^3$, Ehud Altman$^2$\\
% {\small $^1$\em Department of Physics, University of Toronto, Toronto, Ontario M5S 1A7, Canada}\\
% {\small $^2$\em Department of Condensed Matter Physics, Weizmann Institute of Science, Rehovot 7610001, Israel}\\
% {\small $^3$\em Institute of Theoretical Physics, University of Cologne, D-50937 Cologne, Germany}}

\begin{abstract}
  We investigate the superfluid properties of two-dimensional driven Bose
  liquids, such as polariton condensates, using their long-wavelength
  description in terms of a compact Kardar-Parisi-Zhang (KPZ) equation for the
  phase dynamics. We account for topological defects (vortices) in the phase
  field through a duality mapping between the compact KPZ equation and a theory
  of non-linear electrodynamics coupled to charges. Using the dual theory we
  derive renormalization group equations that describe vortex unbinding in these
  media. When the non-equilibirum drive is turned off, the KPZ non-linearity
  $\lambda$ vanishes and the RG flow gives the usual Kosterlitz-Thouless (KT)
  transition. On the other hand, with non-linearity $\lambda>0$ vortices always
  unbind, even if the same system with $\lambda=0$ is superfluid. We predict the
  finite size scaling behavior of the superfluid stiffness in the crossover
  governed by vortex unbinding showing its clear distinction from the scaling
  associated with the KT transition.
\end{abstract}

\maketitle

\section{Introduction}
\label{sec:introduction}

Recent experiments involving strong coupling of matter and light are advancing
quantum optics to the domain of many-body
physics~\cite{Baumann2010,Schauss12,Houck2012,Underwood2012,Britton2012,Blatt2012}. Motivated
by this progress, theoretical efforts are under way to understand emergent
phenomena in driven open quantum
systems~\cite{Lechner2013,Altman2015,Piazza2014,Keeling2014,Brennecke2015,Raftery2014,Hartmann2008,Ritsch2013,DallaTorre2010,Sieberer2013,Marino2016,Sieberer2015a}. One
of the best studied model systems in this class are fluids of exciton-polaritons
in semiconductor microcavities, which have shown condensation-like
phenomena~\cite{Kasprzak2006,Balili2007,Deng2007,Roumpos2012,
  Belykh2013,Nitsche2014}. Because of their photonic component
exciton-polaritons have a finite life-time, which necessitates continuous
pumping with light in order to maintain a steady state. A fundamental question
we address in this paper is whether a fluid subject to such non-equilibrium
conditions can be superfluid.

Aspects of superfluid behavior have been studied extensively in microcavity
polariton systems. Questions that have been addressed theoretically are the
proper generalization of the Landau criterion to driven-dissipative
condensates~\cite{Carusotto2004,Wouters2010b,Wouters2010a,Pigeon2011} in which
the linearly dispersing Bogoliubov sound mode is replaced by a diffusive
mode~\cite{Szymanska2006,Wouters2006}; the superfluid and normal fractions in a
homogeneous driven open condensate were calculated in~\cite{Keeling2011}, and
Refs.~\cite{Janot2013,Gladilin2016} studied, on the mean-field level, the
influence of external potentials on these quantities. Experimentally, features
such as quantized vortices~\cite{Lagoudakis2008}, suppression of scattering off
obstacles~\cite{Amo2009a,Amo2009b,Amo2011,Grosso2011,Nardin2011}, and
metastability of persistent currents~\cite{Sanvitto2010} have been
observed. These studies suggest that driven-dissipative condensates support a
superfluid phase as do their equilibrium counterparts. This is corroborated by
experimental~\cite{Deng2007,Roumpos2012,Nitsche2014} and
numerical~\cite{Dagvadorj2015} investigations of spatial coherence in such
systems, which found a transition from short-range to algebraic order as the
strength of external pumping is increased. Thus, at high pump power
exciton-polariton systems show signatures of quasi-long-range order and
superfluidity analogous to the KT phase realized in Bose liquids in thermal
equilibrium at low temperatures.

\begin{figure}
  \centering  
  \includegraphics[width=.9\linewidth]{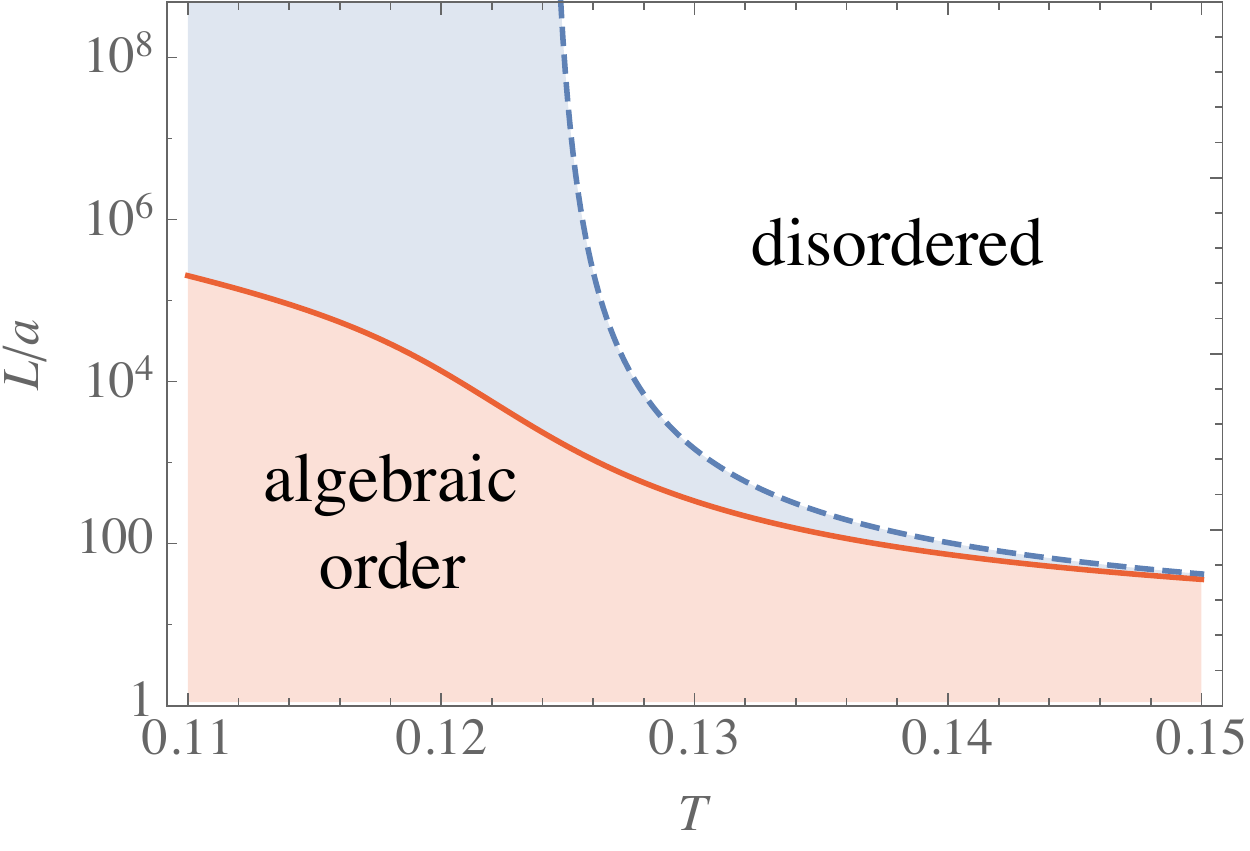}  
  \caption{(Color online.) Finite-size phase diagram for a 2D driven-dissipative
    condensate. $T$ is the ``vortex temperature,'' i.e., the strength of the
    noise acting on topological defects, and $L$ is the system size measured
    here in units of a microscopic cutoff scale $a$. In thermal equilibrium (blue,
    dashed line), close to the critical temperature $T_c$, the phase
    boundary behaves as $\ln(L/a) \sim 1/\sqrt{T_c - T}$. Out of
    equilibrium (red, solid line), algebraic order is destroyed at any noise
    strength in the thermodynamic limit. For small values of $T$, the phase
    boundary approaches a value on the order of the bare screening length
    $L_v$. The phase boundaries are obtained by integrating the renormalization
    group flow equations~\eqref{eq:RGflow} up to $y = 1$, starting from an
    initial value of $y = 0.1$.
  }
  \label{fig:phase_diagram}
\end{figure}

Recently it has been
noted~\cite{Altman2015,Keeling2016,Gladilin2014,Ji2015,He2015} that the long
wavelength fluctuations of a driven dissipative condensate map to a
Kardar-Parisi-Zhang equation~\cite{Kardar1986}
\begin{equation}
  \label{eq:KPZ}
  \partial_t \theta = D \nabla^2 \theta + \frac{\lambda}{2} \left( \nabla\theta
  \right)^2 + \eta,
\end{equation}
where the condensate phase field $\theta(\mathbf{x},t)$ here plays the role of
the height field $h(\mathbf{x},t)$ of the original interface growth problem, and
the non-linearity $\lambda$ is proportional to the deviation from effective
thermal equilibrium~\cite{Altman2015}. The renormalization group (RG) analysis
of the KPZ equation shows that the non-linear term is relevant in two
dimensions, exhibiting a flow to a strong coupling fixed
point~\cite{Canet2010,Canet2011b,Canet2012,Kloss2012} in which the interface is
``rough,'' or more precisely, height correlations increase as a power-law of
distance
$\av{ \left( h(\mathbf{x}, t)-h(0,t) \right)^2} \sim \abs{\mathbf{x}}^{2 \chi}$,
where $\chi >0$ denotes the ``roughness exponent.'' For the driven condensate
problem the rough phase, corresponding to strongly non-linear fluctuations of
the phase, is manifested by a stretched exponential decay of the condensate
correlations. However, these correlations establish only beyond a large emergent
length scale $L_*$, while at shorter distances the correlations can show
algebraic decay as in equilibrium. Thus, the algebraic order observed in
exciton-polaritons can only be a finite-size crossover phenomenon. This raises
the question, whether the same is true for superfluidity: can a
driven-dissipative condensate support a superfluid phase in the thermodynamic
limit or will it ultimately be destroyed by diverging fluctuations?

There is a crucial difference between Eq.~\eqref{eq:KPZ} and the KPZ equation
that is not taken into account by the above considerations and is pertinent to
the question of superfluidity. Contrary to the height field, the phase $\theta$
is compact, defined periodically on the interval $[ 0,2 \pi )$, hence the phase
admits topological defects that the conventional non-compact height field does
not. This difference also arises in ``active smectics''~\cite{Chen2013} and
driven vortex lattices in disordered
superconductors~\cite{Aranson1998}. Similarly, the dynamics of the phase of
sliding density waves~\cite{Balents1995,Chen1996} and arrays of coupled
limit-cycle oscillators~\cite{Lauter2015,Lauter2016} feature KPZ-type
non-linearities. It may be natural to expect that the rapid (i.e., faster than
power-law) decay of condensate correlations would lead to unbinding and
proliferation of vortices rendering the strong coupling fixed point
unstable. However this expectation is based on our understanding of the
unbinding of vortices in the Kosterlitz-Thouless
transition~\cite{Kosterlitz1973,Berezinskii1971,Berezinskii1972}, which is valid
only in thermal equilibrium. On the other hand, we show below by direct analysis
of the strong coupling KPZ fixed point (i.e., neglecting vortices), that such a
state sustains a finite superfluid stiffness that may conceivably protect from
proliferation of unbound vortices even if they were allowed. These seemingly
conflicting considerations concerning the stability of a superfluid steady state
highlight the need to account for the vortices within a comprehensive theory of
the non-equilibrium condensate fluctuations.

In this paper, we incorporate vortices into the framework by establishing a
duality between the compact KPZ equation~\eqref{eq:KPZ} and a non-linear
electrodynamics theory. Our heuristic derivation here is complemented by a more
systematic approach with the same result in~\cite{Sieberer2016}. The question of
superfluidity is thereby translated into the problem of screening of charges
(vortices) by the non-linear medium. As in equilibrium, if unbound charges
proliferate, then they screen the electric field (circulating persistent
current) emanating from a test charge, thus destroying the superfluid properties
of the system.

We solve the non-linear screening problem within a perturbative renormalization
group scheme valid for low vortex density and at length scales below the KPZ
scale $L_*$, where the non-linearity $\lambda$ can be considered small. In the
equilibrium case in which $\lambda = 0$, and if the number of particles is
conserved, our approach reproduces the standard linear electrodynamics of
vortices in superfluid
films~\cite{Ambegaokar1978,Ambegaokar1980,Cote1986,Minnhagen1987} exhibiting a
Kosterlitz-Thouless transition, while out of equilibrium it allows us to
systematically account for the effect of the non-linearity on the dynamics of
vortices. The main effect of the non-linearity is to modify the force law
between pairs of vortices so that they always unbind at length scales larger
than the emergent scale $L_v$. The latter is exponentially large in the
parameter $\lambda/ D$, which we assume to be small for simplicity. Note that
this length scale is different from the KPZ scale $L_*$, which is exponentially
large in $g^2=\lambda^2\Delta/ D^3$. In particular, the KPZ scale $L_*$ depends
on the noise strength, whereas $L_v$ does not.

Indeed we note that the instability towards vortex unbinding has already been
discussed in Ref.~\cite{Aranson1998b} in the context of the complex
Ginzburg-Landau equation and in Ref.~\cite{Aranson1998} for the driven vortex
lattice. This result was based on solving the equations of motion of the
vortices within the deterministic (i.e., noiseless) non-linear equations and in
this way obtaining their mutual interaction. In this paper we come to this
conclusion from a different angle, through derivation of the dual electrodynamic
theory. This allows us not only to identify the instability, but also to predict
and characterize the different crossover regimes using a renormalization group
analysis. In particular, we characterize the crossover from Kosterlitz-Thouless
equilibrium-like physics at intermediate scales to the unbinding governed by the
non-equilibrium physics occurring at longer scales. Using the RG analysis we can
estimate the regimes of parameters where the former physics would be observed in
experiment and where the new emergent effects of the non-equilibrium
fluctuations become dominant. This crossover is seen in the finite size phase
diagram shown in Fig.~\ref{fig:phase_diagram}.
%
%Current experimental
%realizations of driven-dissipative condensates with exciton-polaritons are
%are not probing sufficiently large systems to see a significant effect of the KPZ non-linearity,
%but the parameters of the system can be changed to enhance the there fluctuations~(see \cite{Altman2015,He2015}).

\section{Superfluidity in a vortex-free driven-dissipative system}
\label{sec:sf}

Before discussing the effect of vortices it is interesting to investigate the
fate of superfluidity in a driven-dissipative system without topological
defects. That is, we ask whether a system described by the KPZ
equation~\eqref{eq:KPZ} with a {\em non-compact} phase field has a
superfluid response to an external vector potential. It is well known that
within the standard analysis of the KT transition in the equilibrium $XY$ model
the superfluid stiffness is not renormalized in absence of vortices. However,
the KPZ equation provides an independent source of non-linearity that can
potentially destroy superfluidity even without vortices. This possibility is
perhaps further suggested by the fact that the non-linearity does lead to
destruction of the condensate correlations, leading to (stretched) exponential
rather than power-law decay.

Below we show that this naive expectation is in fact incorrect. In the absence
of topological defects in the phase field, the physics of a 2D
driven-dissipative condensate is governed by the strong-coupling fixed point of
the non-compact KPZ equation.  We show that this implies very peculiar
properties: while correlations of the condensate field are short ranged, the
response is superfluid. We stress again that this is the outcome only if we
neglect the generation and possible proliferation of vortices.

In Ref.~\cite{Keeling2011}, the superfluid density of a 2D driven-dissipative
condensate was calculated using an expansion in fluctuations around the
mean-field stationary state. Such an approach is valid when phase fluctuations
remain small, i.e., when the nonlinear term in the KPZ equation is not dominant
as in finite systems below the KPZ scale $L_{*}$. Also in this case the response
is superfluid, however, the mechanism leading to this result is completely
different from the one that yields the finite superfluid density we obtain
below. We defer a detailed discussion of this point to
Appendix~\ref{sec:sf-density-kpz}.
 
In the following, we work with the effective action that generates the KPZ
equation~\eqref{eq:KPZ},
\begin{equation}
  \label{eq:S_KPZ}
  S_{\mathrm{KPZ}} = \int d^2 \mathbf{r} d t \, \hat{\theta} \left[ \partial_t
    \theta - D \nabla^2 \theta  - \frac{\lambda}{2} \left( \nabla
      \theta\right)^2 - \Delta \hat{\theta} \right].
\end{equation}
The fields $\theta$ and $\hat{\theta}$ are independent variables. In the
Martin-Siggia-Rose (MSR) framework~\cite{Kamenev2011,Altland2010a,Tauber2014a},
$\hat{\theta}$ is termed the response field.

In presence of $U(1)$ symmetry,\footnote{In the MSR framework, due to the
  ``doubling'' of degrees of freedom, one has to distinguish two different types
  of phase rotation transformations. Only one of them, corresponding to a shift
  of $\theta$ by a constant, is a symmetry of the KPZ action~\eqref{eq:S_KPZ},
  while the other one is not a symmetry in the absence of particle number
  conservation~\cite{Sieberer2015a}.} superfluidity can be defined in the driven
open system --- even without particle number conservation --- through the
response of the flow to gauging the $U(1)$ symmetry with an external vector
potential. Formally, this is done through the usual replacement
$\nabla \theta \to \nabla \theta - \mathbf{a}$; physically, the vector potential
$\mathbf{a}$ can be introduced, e.g., in exciton-polaritons using the method
described in Ref.~\cite{Keeling2011}. Expanding the action to linear order in
the applied vector potential we get
$S[{\mathbf a}]=S_{\mathrm{KPZ}} - \int d^2\br dt \,{\mathbf a}\cdot
\hat{\mathbf j}$. The object linearly coupled to the gauge field,
\begin{equation}
  \label{eq:3}
  \hat{\mathbf j} = D \nabla \hat{\theta}- \lambda \hat{\theta} \nabla \theta,
\end{equation}
is the Noether current associated with the $U(1)$ symmetry. It is {\em not} the
physical particle current because the symmetry is not conjugate to particle
number conservation~\cite{Sieberer2015a}. In absence of particle number
conservation it is more subtle to identify the physical current ${\mathbf j}$
that responds to the applied force. This can be done by noting that in the
extended system including also the bath degrees of freedom the number of
particles is conserved. Hence one can define the total current density in this
enlarged system and decompose it into two components: (i) the dissipative
current flowing from the system into the bath, and (ii) the non-dissipative
current which flows only in the system~\cite{Avron2012,Gebauer2004}. In our
case, the non-dissipative current is given by the standard expression
${\mathbf j}=D \nabla\theta$ (see~\cite{Keeling2014} for a proposal of an
experimental procedure to address this current in microcavity polaritons by
means of artificial gauge fields). Now the superfluid stiffness can be defined
in terms of the linear response of this current to the applied vector potential:
\begin{equation}
  \label{eq:4}  
  \chi_{i j}(\mathbf{x} - \mathbf{x}', t - t') = \left. \frac{\delta
      \left\langle j_i(\mathbf{x}, t) \right\rangle}{\delta a_j(\mathbf{x}',
      t')} \right\rvert_{\mathbf{a} = 0}
  = \av{j_i(\mathbf{x}, t) \hat{j}_j(\mathbf{x'}, t')}.
\end{equation}
The response function can be decomposed into two contributions,
$\chi = \chi^{(1)} + \chi^{(2)}$, which are
\begin{align}
  \label{eq:sfd_45}
  \chi_{ij}^{(1)}(\mathbf{x} - \mathbf{x}', t - t')
  & = D^2 \langle \partial_i \theta(\mathbf{x}, t) \partial_j
    \hat{\theta}(\mathbf{x}', t') \rangle,
  \\ \label{eq:sfd_46}
  \chi_{ij}^{(2)}(\mathbf{x} - \mathbf{x}', t - t')
  & = D \lambda \langle \partial_i \theta(\mathbf{x}, t) \partial_j
    \theta(\mathbf{x}', t') \hat{\theta}(\mathbf{x}', t') \rangle.
\end{align}
The superfluid stiffness in an isotropic system is directly related to the
difference between the longitudinal and transverse parts of the Fourier
transform of the response function at zero frequency in the long-wavelength
limit~\cite{Griffin1994,Hohenberg1965,Pitaevskii2003}. (An alternative
definition of the superfluid stiffness in terms of the oscillation frequency of
the condensate was used in Refs.~\cite{Janot2013,Gladilin2016}. We note that
these works did not consider the effect of fluctuations on the superfluid
density.)
  
The contribution $\chi^{(1)}$ to the response function looks like the standard
current-current response in equilibrium superfluids. Indeed, while $\chi^{(1)}$
is scale invariant at the Gaussian fixed point corresponding to an equilibrium
superfluid, the contribution $\chi^{(2)}$ vanishes because it is the average of
an odd number of fields.
% vanishes with the system size as
% $L\to \infty$ (as $1/L$ ?).
However, in the driven-dissipative case, the appropriate steady state (still
ignoring vortices) is governed by the strong coupling fixed point of the KPZ
equation. As we show in Appendix~\ref{sec:sf-density-kpz}, at this fixed point
the contribution $\chi^{(1)}$ has a negative scaling dimension of $-\chi$, with
$\chi$ being the roughness exponent of the strong coupling fixed
point. Numerical
simulations~\cite{Kim1989,Miranda2008,Marinari2000,Ghaisas2006,Chin1999,Tang1992,Ala-Nissila1993,Castellano1999,AaraoReis2004,Kelling2011,Halpin-Healy2012,halpin-healy_universal_2014,Pagnani2015}
and functional RG~\cite{Canet2010,Canet2011b,Canet2012,Kloss2012} results find
$\chi \approx 0.4$. Hence the conventional part of the current-current response
function gives a vanishing contribution to the superfluid stiffness. On the
other hand, $\chi^{(2)}$ is scale invariant at the strong coupling fixed point and
leads to a constant superfluid stiffness.% {\cred{ A more
    % detailed analysis shows that superfluid stiffness of an infinite system
    % assumes a universal form at the fixed point given by
    % $\rho_s= (8\pi)^{-1}\ln 2 g_*\rho_0$ where $g_*$ is the universal value of
    % the coupling constant at the strong coupling fixed point [I think we can
    % omit the the last sentence because the final result is ultimately non
    % universal (a universal dimensionless number times a non-universal constant
    % with dimensions of stiffness). EA]}}.
 
We conclude that in a driven-dissipative condensate governed by KPZ dynamics the
superfluid response is a non vanishing constant in spite of not having any long
range or even algebraic order. However, as we mentioned before, this result
assumes that the phase field behaves as a non-compact variable neglecting the
existence and possible proliferation of topological defects in it. Below we
develop a theory that incorporates the vortices into the general non-equilibrium
framework.

\section{Dual electrodynamic theory}
\label{sec:vortex-dynamics}

The force between point vortices in conventional two-dimensional superfluids
falls off as the inverse distance, exactly as the force between two dimensional
Coulomb charges. Such a duality mapping between vortices and electrostatics was
famously exploited in the theory of the Kosterlitz-Thouless
transition~\cite{Kosterlitz1973,Jose1977,Jose1978,Savit1980}. The transition
from the low-temperature superfluid to the high-temperature normal state is dual
to a two-dimensional Coulomb gas undergoing a transition from bound dipoles to a
plasma of free charges. The superfluid stiffness is, in the dual picture, given
by the inverse dielectric constant of the Coulomb gas, which in the plasma phase
falls to zero at long distances, due to effective screening of the Coulomb
forces.

To investigate the stability of a superfluid in non-equilibrium conditions,
we extend the duality to an electrodynamic theory that takes into account the
dynamics under the influence of particle loss and external drive. In this paper
we provide a simple heuristic derivation working with a continuum theory
throughout. A systematic derivation of the same dual theory on a lattice is
given in a parallel paper~\cite{Sieberer2016}. We note in passing that in the
context of zero-temperature superfluid-insulator transitions the vortex-charge
duality has been extended to a complete quantum-electrodynamics
theory~\cite{Fisher1989a}. In the system we consider, the non-equilibrium
conditions of the condensate lead to a dual description in terms of classical
electrodynamics with crucial modifications.

\subsection{Modified Maxwell equations with charges}
\label{sec:modif-maxw-equat}

The starting point of the duality mapping is the usual low-frequency description
of driven open Bose liquids, such as exciton-polariton systems, the stochastic
Gross-Pitaevski or complex Landau-Ginzburg equation (CGLE) for the complex
scalar order parameter~\cite{Carusotto2013},
\begin{equation}
\partial_t\psi(\br,t) = - {\d H_d\over \d \psi^*} - i {\d H_c\over \d\psi^*}
+\zeta({\bf r},t).
\label{EoM}
\end{equation}
$H_\ell$ ($\ell=c,d$) are effective Hamiltonians, which generate coherent and
dissipative dynamics, respectively, and read for an isotropic situation \be
H_\ell= \int d^2\br \left[r_\ell |\psi|^2+K_\ell |\nabla\psi|^2 +\half u_\ell
  |\psi|^4\right]~.
\label{Hcd}
\ee The term $\zeta(\br,t)$ in Eq.~(\ref{EoM}) describes a Gaussian white noise
with short-ranged spatio-temporal correlations
$\av{\zeta^*(\br,t)\zeta(\br',t')}=2\sigma\d(\br-\br')\d(t-t')~,
\av{\zeta(\br,t)\zeta(\br',t')}=0$,
and zero mean. In the following, we set $K_d=0$ for simplicity, which also is a
good approximation in exciton-polariton systems~\cite{Carusotto2013}. It is
straightforward to include this term and verify that it does not lead to a
different effective theory.

The complex field $\psi(\mathbf{r},t)$ is conveniently split into two real
variables in terms of the phase-amplitude representation, \be \psi
(\mathbf{r},t) = \sqrt{\bar \rho + \delta \! \rho(\mathbf{r},t)}e^{ i \theta
  (\mathbf{r},t)}.  \ee Here, the total density under the square root can be
expanded in the fluctuations $ \delta \! \rho(\mathbf{r},t)$ about the
homogeneous background density $\bar \rho$, since density fluctuations are
damped or gapped at long wavelength. In contrast, the phase fluctuations
describe the Goldstone mode of the phase rotation symmetry at sufficiently low
noise level, and thus are gapless (cf. e.g. the discussion in
\cite{Altman2015}). Therefore, in the coupled stochastic equations of motion for
phase and amplitude fluctuations, it is possible to directly integrate out the
density mode, which leads to the KPZ equation~\eqref{eq:KPZ}---an effective
long-wavelength description for the phase mode, valid at scales well below the
damping gap of the amplitude fluctuations~\cite{Altman2015}. The parameters
appearing in the KPZ equation are related to those of the CGLE as
$D = K_c u_c/u_d$ and $\lambda = - 2 K_c$, and the Gaussian noise field $\eta$
in Eq.~\eqref{eq:KPZ} has white correlations with strength
$\Delta = \sigma (u_c^2 + u_d^2)/(2 u_d \bar \rho)$,
\begin{equation}
  \label{eq:Delta}
\langle \eta(\br,t) \eta(\br',t') \rangle = 2\Delta\delta(\br-\br')\delta(t-t').
\end{equation}
For exciton-polaritons, a more microscopic description can be formulated in
terms of a generalized Gross-Pitaevskii equation for lower polaritons coupled to
an excitonic reservoir~\cite{Wouters2007a}. The relations between the parameters
entering this model and the KPZ parameters are summarized in
Ref.~\cite{Altman2015}.

In the description of long-wavelength fluctuations of driven-dissipative
condensates in terms of the KPZ equation~\eqref{eq:KPZ}, the compact nature of
the phase variable is not manifest. To make it explicit and to introduce the
main qualitative modification of the problem -- the occurrence of vortices -- in
a direct way, we take here a different route based on a noisy electrodynamic
theory dual to the description in terms of density and phase. More precisely, we
first make the usual~\cite{Ambegaokar1980,Fisher1989a} identifications of the
boson density fluctuations with the magnetic field, and the phase gradient with
the electric field: \be \bB = B \bzh = - \delta \! \rho\, \unitvec{z}, \quad
\bE= - \bzh\times\nabla\theta.  \ee The noisy hydrodynamic equations of the
condensate can then be written as equations of electrodynamics in the medium. In
particular, the noisy equation for the phase (density) fluctuations translates
into a modified Amp\`ere (Faraday) law. Most importantly, vortices are
incorporated in this fomulation in a natural way as external charges and
currents for the electromagnetic fields.

We begin with Amp\`ere's law resulting from the equation of motion of the phase. More generally speaking, it describes Euler's equations for the momentum balance in the fluid,
\begin{equation}
  \label{eq:ampere}
  \nabla\times\bB-\frac{\varepsilon}{c}\frac{\partial \bE}{\partial t} =\frac{2\pi}{c} \bJ_m.
\end{equation}
The source term $\bJ_m$ includes both the vortex current density $\bJ_v$ (it turns out below that this term is not crucial, so we do not discuss it in detail here) and a source associated
with the non-equilibrium drive that violates energy and momentum conservation:
\begin{equation}
  \label{eq:10}
  \bJ_m\equiv \bJ_v + \frac{1}{2\pi}\bzh\times\nabla
  \left(\frac{\lambda}{2}E^2+ \eta\right).
\end{equation}
The speed of light takes the value $c = u_c$. We have introduced here a
dielectric constant $\varepsilon$ for consistency, whose physical meaning is
explained below Eq.~\eqref{eq:gauss}.

Now we turn to Faraday's law, associated to the density fluctuations. Indeed, in
a conventional superfluid it stems from the continuity equation for the particle
density. In a driven system with losses, however, we must add a source term to
the continuity equation leading to a corresponding change to Faraday's law as
\begin{equation}
  \label{eq:faraday}
  \nabla\times\bE+\frac{1}{c}\frac{\partial\bB}{\partial t}+\gamma\bB  = 0.
\end{equation}
%{\color{red} The noise correlations for $\bar \eta$ take the same form as discussed above for $\eta$, and $\gamma = 2 u_d\bar \rho/u_c$.} 
Here the dissipative coefficient $\gamma$ must appear from symmetry
considerations. Below we relate it to the more microscopic parameters of the
CGLE model~\eqref{EoM}. For completeness we note that the continuity equation
derived from the CGLE of the driven condensate~\cite{Altman2015} generically
contains also a noise term. We omit it here since ultimately it will just add to
the noise source already present in Eq.~\eqref{eq:ampere}.

Two more equations are missing to complete the analogy to Maxwell's equations. The homogenous Maxwell equation $\nabla\cdot{\bB}=0$ is trivial in the two dimensional setting as $\bB=B\bzh$. The condition of irrotational
flow (in the absence of vortices) translates to Gauss' law for the electric
field (as appropriate for the 2D case, the numerical factor on the RHS is
$2 \pi$ instead of the usual $4 \pi$),
\begin{equation}
  \label{eq:gauss}
  \nabla\cdot\bE={2\pi\over \varepsilon} n_v,
\end{equation}
where the vortex density
$n_v({\mathbf r})=\sum_i n_i \delta({\mathbf r}-{\mathbf r}_i)$ acts as a source
of the electric field. In constrast to the vortex current in Eq.~\eqref{eq:10},
the vortex density will play a crucial role for the understaning ot the
problem. Note the appearance of the dielectric constant $\varepsilon$ as in the
analysis of the KT transition. It anticipates that upon coarse graining the
electric field emanating from a test charge will be screened by fluctuations
consisting of bound vortex pairs with separation smaller than the running cutoff
scale. At the microscopic scale, where the above equations are formulated, it
takes the value $\varepsilon =1$, but then will be substantially renormalized.

Finally, to have a complete dynamical description we must also supplement the equations
for the fields~\eqref{eq:gauss},~\eqref{eq:ampere} and~\eqref{eq:faraday} with
dynamics for the charges (vortices) subject to those fields. In principle the
charges are affected both by the electric and magnetic forces. However,
consistent with the assumption of over-damped dynamics we include only the
electric forces~\cite{Ambegaokar1980}
\begin{equation}
  \label{eq:langevin}
  \frac{\partial \br_i}{\partial t}=\mu n_i \bE(\br_i,t) + \bxi_i,
\end{equation}
where the vortex mobility $\mu$ is a free parameter of the theory, which in
principle can be determined from numerical simulations of the CGLE
equation. Note, however, that in reality the CGLE is not the full microscopic
model of the system, but only an intermediate scale effective theory. Hence it
is better to treat $\mu$ as a new independent variable. Assuming that the
universal long-wavelength behavior does not depend on the value of $\mu$, below
we consider the limit $\mu \ll D$. Then, the motion of vortices is slow as
compared to the fluctuations of the fields, which greatly simplifies the
theoretical analysis of the problem. $\bxi_i$ is a random force field on the
vortices, which we take to be short range and time correlated
\begin{equation}
  \label{eq:temp}
  \langle\xi_{i\alpha}(t)\xi_{j\beta}(t')\rangle=2\mu T\delta_{ij}
  \delta_{\alpha\beta}\delta(t-t').
\end{equation}
Microscopically the effective temperature for the vortices and the phase noise
$\eta$ should be closely related as they have the same origin. However upon
rescaling they may flow independently and therefore we keep them as different
quantities in the effective model. This is to be compared with the equilibrium
case where one must have $T = \Delta / D$ at all scales. We further note that the
dynamical equation~\eqref{eq:langevin} does not allow for creation and
annihilation of vortices, however we will introduce an effective fugacity $y$
which tunes their density in the steady state.

We can easily make contact to the KPZ equation~\eqref{eq:KPZ}. Since we are interested in the low frequency limit, we can neglect
$\dot{\bB}$ compared to $\bB$ in Eq.~(\ref{eq:faraday}): at long scales the density fluctuations are
over-damped. The magnetic field can thus be
eliminated from the equations using the over-damped  Faraday law
$\bB=-\gamma^{-1}  \nabla\times\bE $ .  Plugging this as well as Gauss'
law~\eqref{eq:gauss} into~\eqref{eq:ampere} we obtain a dynamical equation for the electric field $\bE$ alone,
\begin{equation}
  \label{eq:Edyn}
  \varepsilon \frac{\partial\bE}{\partial t} = D\nabla^2\bE -
  \bzh\times\nabla \left(\frac{\lambda}{2}E^2+\eta\right) - {2\pi\over \varepsilon} D\nabla n_v-2\pi\bJ_v.
\end{equation}
where $D\equiv c/\gamma$. Now it is easily verified that by setting the vortex
current and density to zero and re-expressing the field in terms of $\nabla\t$,
we recover Eq.~\eqref{eq:KPZ}.

\subsection{Maxwell action with charges}\label{sec:gauge}

In this section we develop a functional integral formulation of the
generating functional for dynamic observables. We then use this
formulation to derive the effective interactions between the charges
mediated by the fluctuating electromagnetic fields and study whether
vortex anti-vortex pairs tend to unbind.
  
As in usual electrodynamics it is convenient to write the electric and
magnetic fields in terms of gauge potentials which automatically solve
the homogenous Maxwell equations. The relation between the gauge
potentials and the fields is, however, altered due to the over-damped
dynamics. We set:
\begin{equation}
  \label{eq:gauge}
  \bB=\frac{1}{\gamma}\nabla\times\bA,\qquad\qquad \bE=-\nabla\phi-\bA,
\end{equation}
which differs from the usual relation, where $\partial_t \bA$ instead of $\bA$ appears in the definition of the electric field. The dynamical equations are then invariant under the local gauge
transformation
\begin{equation}
  \label{eq:gtr}
  \bA\to\bA+\nabla\chi,\qquad\qquad\phi\to\phi-\chi
\end{equation}
(here, $\chi$ appears instead of the usual $\partial_t \chi$ in the last relation). Note that this local gauge freedom reflects the parameterization of the electric and magnetic fields, and is not directly related to the global $U(1)$ invariance of the original phase degree of freedom. For our purposes it is sufficient to choose the following gauge
\begin{equation}
  \label{eq:lorentzg}
  \frac{\partial\phi}{\partial t}+\frac{D}{\varepsilon}
  \nabla\cdot\bA=0,
\end{equation}
corresponding to the usual Lorentz gauge in electrodynamics.
% Functional integration over fields with a gauge symmetry
% is conveniently done by averaging over a Gaussian distribution of gauge
% choices, akin to the Fadeev-Popov treatment of non-Abelian gauge
% fields in QFT. Thus, we introduce the random gauge choice,
% \begin{equation}
%   \label{eq:randomg}
%   \frac{\partial\phi}{\partial t}+\frac{D}{\varepsilon}
%   \nabla\cdot\bA=\zeta,
% \end{equation}
% where 
% \begin{equation}
%   \label{eq:2}
%   \braket{\zeta(\br,t) \zeta(\br',t')}=2\Delta'\delta(\br-\br')
%   \delta(t-t').
% \end{equation}
The gauge fields then obey the following equations
% \begin{eqnarray}
%   \frac{\partial \phi}{\partial t}-\frac{D}{\varepsilon}\nabla^2\phi
%   & = & \zeta + \frac{2\pi D}{\varepsilon^2}n_v 
%     \label{eq:phi} \\
%   \frac{\partial \bA}{\partial t}-\frac{D}{\varepsilon}\nabla^2\bA
%   & = & -\nabla\zeta+\frac{2\pi}{\varepsilon}\bJ_v-
%  { \bzh\times\nabla\over \varepsilon}\left(\eta+\frac{\lambda}{2}E^2\right)
%   \label{eq:A}
% \end{eqnarray}
\begin{eqnarray}
  \frac{\partial \phi}{\partial t}-\frac{D}{\varepsilon}\nabla^2\phi
  & = & \frac{2\pi D}{\varepsilon^2}n_v ,
    \label{eq:phi} \\
  \frac{\partial \bA}{\partial t}-\frac{D}{\varepsilon}\nabla^2\bA
  & = & \frac{2\pi}{\varepsilon}\bJ_v +
 { \bzh\times\nabla\over \varepsilon}\left(\frac{\lambda}{2}E^2 + \eta \right).
  \label{eq:A}
\end{eqnarray}

Using the standard MSR framework we write a functional integral which
generates the stochastic equations for the gauge fields~\eqref{eq:phi}
and~\eqref{eq:A} and $N$ charges
\begin{eqnarray}
  \label{eq:Z}
  Z(N) = \int\D\hat\phi\D\phi\,\D\hat\bA\D\bA\,\D\hat{\bp}\D\br\,
  e^{-S}.
\end{eqnarray}
Here we have already integrated over the gaussian noise fields $\eta$ and
$\bxi$. The hatted fields are the response fields conjugate to those appearing in the equations of motion, which are introduced along the MSR construction.  The action is a sum of the following contributions
$S=S_0+S_\lambda+S_c+ S_{\mathrm{int}}$. The quadratic free field action is given by
\begin{equation}
  \label{eq:S0}
  S_0 = -\frac{1}{2}\int_{\mathbf{r}, t} \begin{array}{cc}
    (\hat A_\mu, & A_\mu) \\ \end{array}
  \left(\begin{array}{cc}
    2\Delta_{\mu\nu} & \delta_{\mu\nu}L^+ \\ \delta_{\mu\nu}L^- & 0
  \end{array}\right)\left(\begin{array}{c}\hat A_\nu \\ A_\nu \end{array}
    \right),
\end{equation}
where $\int_{\mathbf{r}, t} = \int d^2 \mathbf{r} dt$,
$(\hat A_\mu, A_\mu)=(\hat\phi,\hat\bA,\phi,\bA)$,~
$L^\pm = \pm\partial_t-(D/\varepsilon)\nabla^2$, and
\begin{equation}
  \label{eq:8}
  \Delta_{\mu\nu} = \Delta \left( 1-\delta_{\mu 0} \right) \left( 1-\delta_{\nu
      0} \right) \left(\nabla^2\delta_{\mu\nu}
    -\partial_\mu\partial_\nu\right).
\end{equation}

The non-linear contribution from the $\lambda$ term in the KPZ
equation is given by
\begin{equation}  
  S_\lambda = -{\lambda\over 2\varepsilon}\int_{\mathbf{r}, t} \bzh\cdot\left(
    \hat\bA\times\nabla(E^2)\right),
\end{equation}
where $E^2 \equiv (\nabla\phi+\bA)^2$. The combined electromagnetic
action $S_{em}=S_0+S_\lambda$, without coupling to the charges,
provides an equivalent description of the standard non-compact KPZ
equation in two dimensions. This is evident by noting that
$\bv=\bzh\times\bE$ is equivalent to the vector field in Burgers'
equation~\cite{Kardar1986}.

The action of the charges is
\begin{equation}
  \label{eq:9}
  S_c=-\sum_{i=1}^N\int dt\left(\hat{\bp}_i\cdot\partial_t\br_i+\mu T
    \hat{p}_i^2 \right),
\end{equation}
while the coupling of charges to the gauge fields is given by
\begin{equation}
  S_{\mathrm{int}}=\int_{\mathbf{r}, t} \left( \frac{2 \pi}{\varepsilon}
    \,J_\mu\hat{A}_{\mu}-\mu\,\hat{P}_{\mu} A_{\mu} \right).
\end{equation}
In the above expression for the coupling action we have made use of
the following notations:
\begin{align}
  \label{eq:12}
  J_\mu&\equiv  \left({D\over\varepsilon}\, n_v,\bJ_v\right), \\
\hat{P}_\mu&\equiv(-\nabla\cdot\hat{\bP},\hat{\bP}),
\end{align}
where
\begin{align}
  \label{eq:defs1}
  n_v(\br,t) & = \sum_in_i \delta(\br-\br_i(t)), \\ 
  \bJ_v(\br,t) & = \sum_in_i\partial_t\br_i \delta(\br-\br_i(t)), \\
  \hat{\bP}(\br,t) &= \sum_in_i\hat{\bp}_i \delta(\br-\br_i(t)).
\end{align}

Having derived a complete action describing dissipative non-linear
electrodynamics (equivalent to KPZ dynamics) coupled to charges
(vortices) we can deduce the effective interactions between the
charges in certain limiting cases.

\subsection{Charge interactions for weak nonlinearity}
\label{sec:Coulomb-small-lambda}

In the limit of small $\lambda$ we can integrate over the photon
fields perturbatively in the non-linear coupling with respect to the
quadratic photon action $S_0$. For $\lambda=0$ the effective action
for the charges can be obtained exactly by direct integration over the
gauge fields
\begin{eqnarray}
  \label{eq:Sint2}
  S_I^{(0)}& =& \frac{1}{2}\int_{\br,t;\br',t'}
  \begin{pmatrix}
    2\pi J_\mu/\varepsilon , & -\mu \hat P_\mu
  \end{pmatrix}_{\br,t}\times\nonumber \\ & &
  \begin{pmatrix}
    0 & \delta_{\mu\nu}g^- \\ \delta_{\mu\nu}g^+ & -2 g^+ \Delta_{\mu\nu} g^-
  \end{pmatrix}_{\br,t;\br',t'}
  \begin{pmatrix}
    2\pi J_\nu/\varepsilon \\ -\mu \hat P_\nu
  \end{pmatrix}_{\br',t'},
\end{eqnarray}
where $g^\pm=(L^\pm)^{-1}$. 

Noting that vortices move very slowly in the $\mu/D\to 0$ limit, we find that
the most relevant contribution comes from the static $J_0 P_0$ term, which is
linear in $\mu$, while all other terms are of higher order. Let us denote the
static free Green's function as
\begin{equation}
  \label{eq:G}
  G(\mathbf{r} - \mathbf{r}') = -\nabla^{-2} \delta(\mathbf{r} - \mathbf{r}')
  \simeq - \frac{1}{2 \pi} \ln(|\br-\br'|/a).
\end{equation}
This gives
\begin{equation}
  \label{eq:Slead}
  \begin{split}
    S_I^{(0)} & \simeq 
    % \int_{\mathbf{r}, \mathbf{r}', t} \mu\nabla\cdot
    % \bP(\br,t)\, G(\br-\br')\frac{2\pi  }{\varepsilon}n_v(\br',t) \\
    % & = -\int_{\mathbf{r}, \mathbf{r}', t} \mu\bP(\br,t)\cdot\nabla
    % G(\br-\br')\frac{2\pi}{\varepsilon}n_v(\br',t) \\
    % & = 
    -\int_t \sum_{ij} \hat\bp_i(t)\cdot \mu n_i\nabla_iG(\br_i-\br_j)\frac{2\pi
      n_j}{\varepsilon},
  \end{split}
\end{equation}
which, as expected, describes the 2D Coulomb gas interactions. To see
this we note that the action for a Brownian particle contains the term
$\int dt\, \mu\hat\bp\cdot{\bf f}_0$, where $\bf f_0$ is the force acting
on the particle.  For further reference we note here that, when this
happens to be a conserving force, ${\bf f}_0 = -\nabla V$, the
stationary solution of the corresponding Fokker-Planck equation is just
the Gibbs distribution, $P(\br)\sim e^{-V(\br)/T}$.

Our goal is to calculate the effective mutual forces between two
opposite charges, perturbatively in $\lambda$, in order to determine
the fate of the Kosterlitz-Thouless transition in the presence of the
non-linear term. This task is greatly simplified in the limit $\mu/D
\to 0$, since the fields' response is instantaneous on the time scales
of the vortices' motion. Thus, there are two stages in the calculation
of the forces to a given order $O(\lambda^m)$: (i)
calculation of the $O(\lambda^m)$ terms in the effective
charge action, $S_I^{(m)}$. (ii) taking the limit $\mu/D\to 0$, and
identifying the force as the coefficient of $\hat\bp_i$ linear in $\mu$.

% In the following we calculate corrections to the force from
% perturbations in the nonlinear $\lambda$ term. These corrections can
% be divided into two categories. The first are tree-level corrections,
% which do not include noise correlations, and can be (in principle)
% calculated exactly. The second type of corrections arise from noise
% correlations, and can be treated only within a Renormalization Group
% framework, due to infra-red divergences. We focus on the mutual forces
% within a single vortex-anti-vortex dipole, in order to determine the
% fate of the Kosterlitz-Thouless transition in the presence of the
% non-linear term. This task is greatly simplified in the limit $\mu/D
% \to 0$, since the fields' response is instantaneous on the time scales
% of the vortices' motion.

In what follows we consider the interaction between a vortex ($n_+ =
1$) at $\mathbf{r}_+$ and an antivortex ($n_- = - 1$) at
$\mathbf{r}_-$.
% The correction to the Coulomb law linear in $\lambda$ is represented
% by the tree level diagrams in Fig.~\ref{fig:diag}(b,c).
We begin by determining the correction to the Coulomb law linear in
$\lambda$. The correction to the action contains one term
(corresponding to a tree level diagram)
\begin{widetext}
  \begin{equation}
    \label{eq:TL1a}
    S_I^{(1)} = \frac{2\pi^2\lambda\mu\bzh}{\varepsilon^3}\cdot\int_{1234}
    \hat\bP(\br_1,t_1) g^+(\br_1-\br_2,t_1-t_2)\times\bnabla_2\left(
      g^+(\br_2-\br_3,t_2-t_3)J_\mu(\br_3,t_3)g^+(\br_2-\br_4,t_2-t_4)
      J_\mu(\br_4,t_4)\right),
  \end{equation}
  where $\int_{1\dots M}=\prod_{i=1}^M\int_{\br_i,t_i}$. Note that this term
  is second order in $J_\mu$, demonstrating the breakdown of force superposition
  for $\lambda\ne 0$. In the limit $\mu/D\to 0$, only static
  $J_\mu(\br,t)=(J_0(\br), 0)$ configurations contribute to the MSR integral,
  giving
  % \begin{widetext}
  %   \begin{equation}
  %     \label{eq:TL1}
  %     \begin{split}
  %       S_1 & = -\frac{\lambda}{2D}\left({2\pi\over \varepsilon}\right)^2 \int
  %       dt\,d^2r\mu\bp_0(t)\cdot[\bzh\times{\bf f}_0(\br)] \,\left(\left|{\bf
  %           f}_0(\br-{\bf R})\right|^2 n_0 n_1^2 + 2{\bf
  %         f}_0(\br)\cdot{\bf f}_0({\bf R}-\br)\, n_0^2 n_1\right) \\
  %       & \equiv -\int dt\,\mu \bp_0(t)\cdot{\bf f}_1({\bf R},n_0,n_1)
  %     \end{split}
  %   \end{equation}
  \begin{equation}
    \label{eq:TL1}
    S_I^{(1)} \simeq - \frac{\varepsilon \lambda}{4 \pi D} 
    \int_{\mathbf{r}, t} 
    \mu \hat{\mathbf{p}}_+\cdot \left( \bzh\times{\bf f}_0
      (\mathbf{r}_+ - \mathbf{r}) \right)\left|{\bf f}_0(\br- \mathbf{r}_+) 
      - \mathbf{f}_0(\mathbf{r} - \mathbf{r}_-)
    \right|^2 - (+\leftrightarrow -) 
    \equiv \int_t \mu \hat{\mathbf{p}}_+ \cdot \mathbf{f}^{(1)}
    (\mathbf{r}_+ - \mathbf{r}_-) - (+\leftrightarrow -).
  \end{equation}
\end{widetext}
For simplicity we have denoted,
$\mathbf{f}_0(\mathbf{r}) \equiv \left( 2 \pi/\varepsilon \right) \nabla
G(\mathbf{r}) = -\mathbf{r}/(\varepsilon r^2)$.
The $O(\lambda)$ correction to the force for
$\abs{\mathbf{r}_+ - \mathbf{r}_-}/a \gg 1$ is given by (see
Appendix~\ref{sec:inter-vort-antiv})
\begin{equation}
  \label{eq:1}
  \mathbf{f}^{(1)}(\mathbf{R}) = - \frac{\lambda}{2 D
    \varepsilon^2} \frac{\unitvec{z} \times \mathbf{R}}{R^2} \left( 2 \ln(R/a) - 
    \frac{1}{2} \right).
\end{equation}
This force acts on the dipole in a direction perpendicular to the
segment $\mathbf{R} = \mathbf{r}_+ - \mathbf{r}_-$. Hence, the first
order correction does not affect the probability distribution of the
vortex-dipole sizes.

We next calculate the $O(\lambda^2)$ correction to the
force. There are two types of corrections to the effective action,
$S_I^{(2)}$, at this order. The first involves fluctuation correlations
(corresponding to loop diagrams), and modifies the $J_\mu\hat P_\nu$
and $\hat P_\mu\hat P_\nu$ terms, which appear in the zeroth-order
charge action, Eq.~\eqref{eq:Sint2}. These can be treated only within
a Renormalization Group framework, due to infrared
divergences. Nevertheless, they do not modify the effective force: the
modified $J_\mu\hat P_\nu$ term vanishes in two
dimensions~\cite{Kardar1986}, and the $\hat P_\mu\hat P_\nu$ term is
of order $O(\mu^2)$, which we neglect in the $\mu/D\to 0$
limit. The second type (corresponding to tree level diagrams) is third
order in $J_\mu$, which, in the limit $\mu/D\to 0$ becomes
\begin{widetext}
% \begin{equation}
%   \label{eq:TL2}
%   \begin{split}
%     S_2 &= -\left(\frac{\lambda }{2D}\right)^2\left({2\pi\over
%         \varepsilon}\right)^3\sum_{i,j,k=\pm} \int _{t,\br,\br'} \mu n_0
%     \hat{\mathbf{p}}_+(t)\cdot \left[\bzh\times{\bf f}_0(\br-\br_0)\right]
%     n_i{\bf f}_0(\br-\br_i)\cdot\left[\bzh\times{\bf
%         f}_0(\br'-\br)\right]n_j{\bf f}_0(\br'-\br_j)\cdot{\bf
%       f}_0(\br'-\br_k)n_k \\
%     & \equiv -\int dt\,\mu \hat{\mathbf{p}}_+(t)\cdot{\bf f}_2({\bf r}_1-{\bf r}_0,n_0,n_1)
%   \end{split}
% \end{equation}
  % \begin{equation}
  %   \label{eq:TL2a}
  %   S_I^{(2)} = -2\left(\frac{\lambda}{2\varepsilon}\right)^2\int_{123456}
  %   \mu\bzh\cdot\left[\hat\bP(\br_1,t_1)g^+(\br_1-\br_2,t_1-t_2)\times
  %     \bnabla_2\right(\bnabla
  % \end{equation}
  \begin{equation}
  \label{eq:TL2}
  \begin{split}
    S_I^{(2)} & \sim -\frac{\lambda^2\varepsilon^2}{16\pi^2D^2} \sum_{i,j,k =
      \pm} \int _{\br,\br', t} \mu \hat{\mathbf{p}}_+ \cdot \left( \bzh \times{\bf
        f}_0(\br_+-\br) \right) n_i {\bf f}_0(\br-\br_i) \cdot \left(
      \bzh\times{\bf f}_0(\br-\br') \right) n_j {\bf f}_0(\br'-\br_j)\cdot
    n_k{\bf
      f}_0(\br'-\br_k) - (+\leftrightarrow -) \\
    & \equiv \int dt\,\mu \hat{\mathbf{p}}_+ \cdot \mathbf{f}^{(2)}({\bf r}_+ - {\bf r}_-) +
    (+\leftrightarrow -).
  \end{split}
\end{equation}
\end{widetext}
The correction to the force between two particles of opposite charge and
distance $R\gg a$ is evaluated in Appendix~\ref{sec:inter-vort-antiv}, with the
result
\begin{equation}
  \label{eq:5}
  \mathbf{f}^{(2)}(\mathbf{R}) = \frac{1}{8} \left( \frac{\lambda}{2 D} \right)^2
  \frac{1}{\varepsilon^3} \frac{\mathbf{R}}{R^2} \left( 8 \ln(R/a)^2 + 4
    \ln(R/a) - 1 \right).
\end{equation}
Note that this is a central force that constitutes a repulsive contribution to
the attractive $1/R$ Coulomb law. The last term merely renormalizes the bare
value of the dielectric constant $\varepsilon$. We omit it in the following.

% Finally, the one loop correction to the Coulomb interaction due to
% fluctuations of $\bf A$ (diagram (h) in Fig.~\ref{fig:diag}) vanishes
% identically in two dimensions. Hence ${\bf f}_2$ is the leading
% correction to the Coulomb force in the limit $\mu/D\to 0$.

The effective interactions computed above (see
Appendix~\ref{sec:inter-vort-antiv}) hold as long as the
perturbation theory in $\lambda$ is valid. From this result it is clear
that the second order correction becomes larger than the zeroth order
term beyond a separation of
% $L_v\approx a\exp\left[D^2/\lambda^2C'\right]$
\begin{equation}\label{eq:lv}
L_v = a\, e^{ \frac{2 D}{\lambda}},
\end{equation} 
where we took the bare value of $\varepsilon$ to be $1$.\footnote{\label{fussnote}The bare
  screening length found in Ref.~\cite{Aranson1998} from the solution of a
  single vortex in the KPZ equation is $L_v = a \exp(\pi D/\lambda)$. It differs
  from Eq.~\eqref{eq:lv} in the numerical prefactor in the exponential. This is
  not in conflict with our estimate that is taken from comparing only the
  magnitude of the first and second terms in a perturbative expansion. Defining
  the crossover where the ratio between these two terms is $(\pi/2)^2$ rather
  than 1 would lead to the different numerical factor in the exponent.}
Furthermore perturbation theory in $\lambda$ fails beyond the emergent KPZ
length scale $L_{*} = a \exp(8 \pi D^3/(\Delta \lambda^2))$.
% Note that in the weak noise regime $\D<D$, where the equilibrium system is
% expected to be in the bound dipole phase (superfluid),
% the bare "vortex unbinding" scale $L_v$ is smaller than $L_{*}$.
% \lsc{how do these results compare to Ref.~\cite{Aranson1993}? out of curiosity:
%   could we get the exponential decay at large distances from a resummation of
%   the perturbative series?} {\cred{[Lukas.]}}
Before proceeding we comment on how the corrections to the force,
found above, are related to the flow field around vortices. Consider
first a single vortex solution to the stationary KPZ equation
$D\nabla\cdot\bv+ (\lambda/2) \bv^2=0$, with $\bv=\nabla\t$. It is
easily seen that the velocity field with a vortex in the azimuthal
component $v_\varphi=1/r$ produces a radial component $v_r(r)\sim
-(\lambda/2r)\ln r$ at large $r$. This structure of a vortex is well
known in the literature on complex Ginzburg-Landau equations, where it
is termed a Zhabotinsky spiral~\cite{Aranson2002}. The first order
correction to the Coulomb force is a direct result of this radial flow
- recall that the electric field is ${\bf E}=-\bzh\times {\bf v}$.
Hence the azimuthal flow $v_\varphi$ leads to the usual Coulomb force,
while the radial flow leads to a force perpendicular to it.
% {\cred{[Is
%     there a simple way to explain why that force is not a constant
%     independent of the separation?]}}. 

The second order correction is due to the interaction between the flow fields of
the two vortices. It cannot be explained based on the flow around a single
vortex because the two flows (phases) do not superpose in this non-linear flow
equation. Together with $\mathbf{f}_0$ and in contrast to $\mathbf{f}^{(1)}$, it
contributes to the central part of the force between the vortex-antivortex pair,
$\mathbf{f}_c (\br) = \mathbf{f}_0 (\br) + \mathbf{f}^{(2)} (\br)$, where $\br$
is the relative coordinate between the vortices. This force is again
conservative, and we have the effective potential
$V_{\mathrm{eff}} = - \nabla \mathbf{f}_c$ including an external electric field,
% \begin{equation}
%   \label{eq:Veff}
%   V_{eff}(\br, \bE) \approx \frac{1}{\varepsilon}\ln\frac{|\br|}{a}
%   -\bE\cdot\br-\frac{C\lambda^2}{\varepsilon^3D^2}\ln^2\frac{|\br|}{a}.
% \end{equation}
% $C$ is a numerical constant, and
\begin{multline}
  \label{eq:Veff}
  V_{\mathrm{eff}}(\mathbf{r}, \mathbf{E}) \approx \frac{1}{\varepsilon}
  \ln(r/a) - \mathbf{E} \cdot \mathbf{r} \\ - \frac{\lambda^2}{12 \varepsilon^3
    D^2} \left( \ln(r/a)^3 + c \ln(r/a)^2 \right),
\end{multline}
where $D$ is the diffusion coefficient from the original KPZ
equation. Anticipating independent renormalization of the terms in the
potential, we include a coefficient $c$ with bare value $c_0 = 3/4$. Because
there is an effective potential, the probability for finding the pair separated
by $\bf r$ follows a Gibbs distribution:
% \be
% P({\bf r})= y^2 e^{-V_{\text{eff}}(\br,\bE)/T}\approx 
%   y^2 \left|\frac{\br}{a}\right|^{-\frac{1}{\varepsilon T}}
%   e^{\frac{C\lambda^2}{\varepsilon^3D^2T}\ln^2\frac{|\br|}{a}}
%   \left(1+\frac{1}{T}\bE\cdot\br\right).
%   \ee
\begin{equation}
  \label{eq:Pr}
  P(\mathbf{r}, \mathbf{E}) = y^2 e^{-V_{\mathrm{eff}}(\mathbf{r},\mathbf{E})/T}.
\end{equation}
Here we defined $y^2$, the vortex fugacity, as the probability of having a pair
with a separation of the value of the the short distance cutoff $a$. The
existence of a vortex ``temperature'' $T$ is due to the conservative nature of
the force between them. However, we emphasize again that due to the absence of
global detailed balance, this temperature does not have to coincide with the
noise level of the spin waves.

% \subsection{Effective charge interactions in the strong coupling KPZ fixed point}
% \label{eq:Coulomb-kpz}

\section{Vortex unbinding at small $\lambda$}
\label{sec:KT}

Having obtained the effective interactions between vortices we are ready to
address the central question of this paper concerning the superfluid properties
of the driven condensate. A crucial observation pertinent to this issue is that
the (inverse) static dielectric constant defined in Eq.~\eqref{eq:gauss} is a
direct measure of the superfluid stiffness, $\rho_s$, as defined in
Eq.~\eqref{eq:sfd_4}. % Sec.~\ref{sec:sf}.
To see this, note that fixing a vortex at the origin is equivalent to gauging
the flow with a transverse vector potential ${\bf a}= \bzh\times\hat{\bf r}/R$
(this is a vector potential that couples to the particle currents, not the dual
vector potential coupled to the vortex currents). With no other vortices present
the superfluid response, i.e., the transverse current induced by the vector
potential, can be deduced directly from Gauss' law.  Noting the duality
correspondence $\nabla\theta=\bzh\times\bE$ and its relation to the particle
current ${\bf j}=D\nabla\theta$, we have
${\bf j} = D\bzh\times\bE=D\bzh\times\hat{\bf r}/{\varepsilon R}
=(D/\varepsilon){\bf a}\equiv\rho_s{\bf a}$,
where the electric field due to a vortex at the origin is
$\bE=\hat{\bf r}/\varepsilon R$.  In the presence of noise, on the other hand,
more pairs of charges can be generated and potentially screen the test charge
placed at the origin. This screening, leading to a renormalized reduced value of
$\varepsilon^{-1}$, is the dual description of the renormalized superfluid
stiffness.

The question of superfluidity is thus mapped, as in the coventional
Kosterlitz-Thouless transition, to the problem of screening in the
Coulomb gas, albeit now with interactions modified by the KPZ
nonlinearity. We consider the limit of zero vortex mobility $\mu/D\to
0$, where the vortex interactions are described by a static effective
potential, derived in Sec.~\ref{sec:Coulomb-small-lambda} above in the
limit of small $\lambda$.  The dielectric constant is computed in this
regime using a renormalization group scheme essentially identical to
the one used to describe the KT transition~\cite{Jose1977,Jose1978}.

We begin by calculating the linear response of the ensemble averaged polarization
density to an applied electric field, $\av{\bf \Pi} = \chi
\mathbf{E}$.
% The charge susceptibility $\chi$ is related to the (fully renormalized)
% dielectric constant through $\varepsilon_R=\e+2\pi\chi$.
It can be computed based on the effective Gibbs distribution for the
vortex-antivortex potential Eq.~\eqref{eq:Pr}; to linear order in the
external field $\mathbf{E}$, it is given by
% \begin{eqnarray}
%   \label{eq:pol}
%   \braket{\bf \Pi} & = & \frac{1}{L^2}\int\frac{d^2R}{a^2}
%   \frac{d^2r}{a^2}\,\br\,y^2
%   \left|\frac{\br}{a}\right|^{-\frac{1}{\varepsilon T}}
%   e^{\frac{C\lambda^2}{\varepsilon^3D^2T}\ln^2\frac{|\br|}{a}}
%   \left(1+\frac{1}{T}\bE\cdot\br\right)
%   %% \left(1+\frac{C\lambda^2}{\varepsilon^3D^2T}\ln\frac{|\br|}{a}\right)
%   \nonumber \\ & = & \bE\,\frac{\pi}{T}y^2\int_a^\infty \frac{dr}{a}\left(
%     \frac{r}{a}\right)^{3-\frac{1}{\varepsilon T}}
%   e^{\frac{C\lambda^2}{\varepsilon^3D^2T}\ln^2\frac{r}{a}}\equiv \chi\bE
%   %% \left(1+\frac{C\lambda^2}{\varepsilon^3D^2T}\ln\frac{r}{a}\right).
% \end{eqnarray}
\begin{equation}
  \label{eq:pol}
  \begin{split}
    \langle \bf{\Pi} \rangle & = \frac{1}{L^2} \int \frac{d^2 \mathbf{R}}{a^2}
    \frac{d^2 \mathbf{r}}{a^2} \mathbf{r} P(\mathbf{r}, 0) \left( 1 +
      \frac{\mathbf{E} \cdot \mathbf{r}}{T} \right) \\ & = \frac{\pi y^2}{T}
    \int_a^\infty \frac{dr}{a} \left( \frac{r}{a}\right)^{3-\frac{1}{\varepsilon
        T}} e^{\frac{\lambda^2}{12 \varepsilon^3 D^2 T} \left( \ln(r/a)^3 + c
        \ln(r/a)^2 \right)} \mathbf{E}\\ & = \chi \mathbf{E}.
  \end{split}  
\end{equation}
 We have $1/L^2$ in front, since we are considering the contribution of
{\it one} dipole to the polarization {\it density}. The two area
integrals with measure $1/a^2$ act as the sum over all configuration
of the single dipole, weighted by the probability, Eq.~\eqref{eq:Pr}, and
from which we read off the susceptibility:
% \begin{eqnarray}
%   \label{eq:epsilonR}
%   \varepsilon_R & = & \varepsilon+2\pi\chi \nonumber \\
%   & = & \varepsilon+\frac{2\pi^2}{T}y^2\int_a^\infty \frac{dr}{a}\left(
%     \frac{r}{a}\right)^{3-\frac{1}{\varepsilon T}}
%   e^{\frac{C\lambda^2}{\varepsilon^3D^2T}\ln^2\frac{r}{a}}
%   %% \left(1+\frac{C\lambda^2}{\varepsilon^3D^2T}\ln\frac{r}{a}\right).
% \end{eqnarray}
\begin{equation}
  \label{eq:epsilonR}
  \begin{split}
    \chi = \frac{\pi y^2}{T} \int_a^\infty \frac{dr}{a} \left(
      \frac{r}{a}\right)^{3-\frac{1}{\varepsilon T}} e^{\frac{\lambda^2}{12
        \varepsilon^3 D^2 T} \left( \ln(r/a)^3 + c \ln(r/a)^2 \right)}.
  \end{split}
\end{equation}
The fully renormalized dielectric constant at the largest scales is given
by $\varepsilon_R = \varepsilon + 2 \pi \chi$. As in the conventional Coulomb
gas, this equation is solved using the renormalization group by gradually
increasing the short distance cutoff at infinitesimal steps from $a$ to
$a \left( 1 + d \ell \right)$. Separating the above integral into two parts
\begin{equation}
  \label{eq:intRG}
  \int_a^\infty=\int_a^{a \left( 1 + d \ell \right)} + \int_{a \left( 1+d\ell \right)}^\infty,
\end{equation}
allows to define the dielectric constant at the new cutoff
\begin{equation}
  \label{eq:epsilonp}
  \varepsilon'=\varepsilon+\frac{2\pi^2 y^2}{T} d\ell.
\end{equation}

A renormalization of the other couplings is required to bring the expression for
$\varepsilon_R$ to its original form, Eq.~\eqref{eq:epsilonR},
\begin{widetext}
  \begin{equation}
  \label{eq:redef}
  \begin{split}
    \varepsilon_R%  & = \varepsilon' + \frac{2 \pi^2 y^2}{T} y^2 \int_{a e^{d
      %   \ell}}^\infty \frac{dr}{a} \left(
      % \frac{r}{a}\right)^{3-\frac{1}{\varepsilon T}} e^{\frac{\lambda^2}{12
      % \varepsilon^3 D^2 T} \ln(r/a)^3} \\
    & \approx \varepsilon' + \frac{2\pi^2 y^2}{T} \left[ 1 + \left( 4 -
        \frac{1}{\varepsilon T}\right) d \ell \right] \int_a^\infty \frac{dr}{a}
    \left(\frac{r}{a}\right)^{3-\frac{1}{\varepsilon T} \left( 1 - \frac{c
          \lambda^2}{6 \varepsilon^2 D^2} d \ell \right)} e^{\frac{\lambda^2}{12
        \varepsilon^3 D^2 T} \left[ \ln(r/a)^3 + \left( c + 3 d \ell \right)
        \ln(r/a)^2 \right]} \\ & = \varepsilon' + \frac{2\pi^2 y^{\prime 2}}{T'}
    \int_a^\infty \frac{dr}{a} \left(\frac{r}{a}\right)^{3-\frac{1}{\varepsilon
        T'}} e^{\frac{\lambda^2}{12 \varepsilon^3 D^2 T} \left[ \ln(r/a)^3 + c'
        \ln(r/a)^2 \right]}.
  \end{split}
\end{equation}
\end{widetext}
To obtain the last line we have the following rescaling constraints
\begin{equation}
  \label{eq:rescale}
  \begin{split}
    \frac{y'^2}{T'} & = \frac{y^2}{T} \left[ 1 + \left( 4 - \frac{1}{\varepsilon
          T}\right) d \ell \right] ,\\
    \frac{1}{T'} & = \frac{1}{T} \left( 1 - \frac{c \lambda^2}{6 \varepsilon^2
        D^2} d \ell \right), \\ c' & = c + 3 d \ell.
  \end{split}
\end{equation}
In these RG equations we neglect corrections of order $y^2$ and $\lambda^4$ to
the effective dipole energy. The differential RG equation for $c$ can be solved
immediately and yields $c = c_0 + 3 \ell$ with the bare value $c_0 =
3/4$.
Inserting this result in the flow equations for the remaining couplings, we
obtain

% \begin{eqnarray}
%   \label{eq:RGflow}
%   % T\frac{d\varepsilon}{d\ell} & = & 2\pi^2y^2 \\
%   % \frac{dy}{d\ell} & = & \left(2-\frac{1}{2\varepsilon T}
%   %   \left(1-\frac{C\lambda^2}{\varepsilon^2D^2}\right)\right)y \\
%   % \frac{d\lambda}{d\ell} & = & -\frac{C\lambda^3}{2\varepsilon^3D^2T}.
%   \frac{d\varepsilon }{d\ell} & = &  2\pi^2y^2 \\ 
%   % T\frac{d\varepsilon}{d\ell} & = & 2\pi^2y^2 \\
%   \frac{dy}{d\ell} & = & \left(2-\frac{1}{2\varepsilon T}
%     +\frac{C\lambda^2}{\varepsilon^2D^2}\right)y \\
%     \frac{dT}{d\ell} &=& \frac{2C\lambda^2}{\varepsilon^2 D^2} T
%   % \frac{dT}{d\ell} & = & \frac{2C\lambda^2T}{\varepsilon^2D^2} 
%   %% \\  \frac{d\lambda}{d\ell} & = & \frac{C\lambda^3}{\varepsilon^2D^2}
% \end{eqnarray}
\begin{equation}
  \label{eq:RGflow}
  \begin{split}
    \frac{d \varepsilon}{d \ell} & = \frac{2 \pi^2 y^2}{T}, \\
    \frac{d y}{d \ell} & = \left[ 2 - \frac{1}{2 \varepsilon T} +
      \frac{\lambda^2}{4 \varepsilon^2 D^2} \left( \frac{1}{4} + \ell \right)
    \right] y, \\
    \frac{d T}{d \ell} & = \frac{\lambda^2 T}{2 \varepsilon^2 D^2} \left(
      \frac{1}{4} + \ell \right).
  \end{split}
\end{equation}
Not surprisingly these flow equations are very similar to the KT scaling
equations and collapse to the latter for $\lambda=0$. The effect of the
non-linearity is to always drive the system to the high temperature (plasma)
phase in which charges are perfectly screened.  However this screening is
achieved only at large length scales, bounded from below by the bare vortex
unbinding length $L_v = a \exp(2
D/\lambda)$. % $L_v = a \exp\left[D^2/(C\lambda^2)\right]$.

\begin{figure}[t]
  \centering  
  \includegraphics[width=0.9\linewidth]{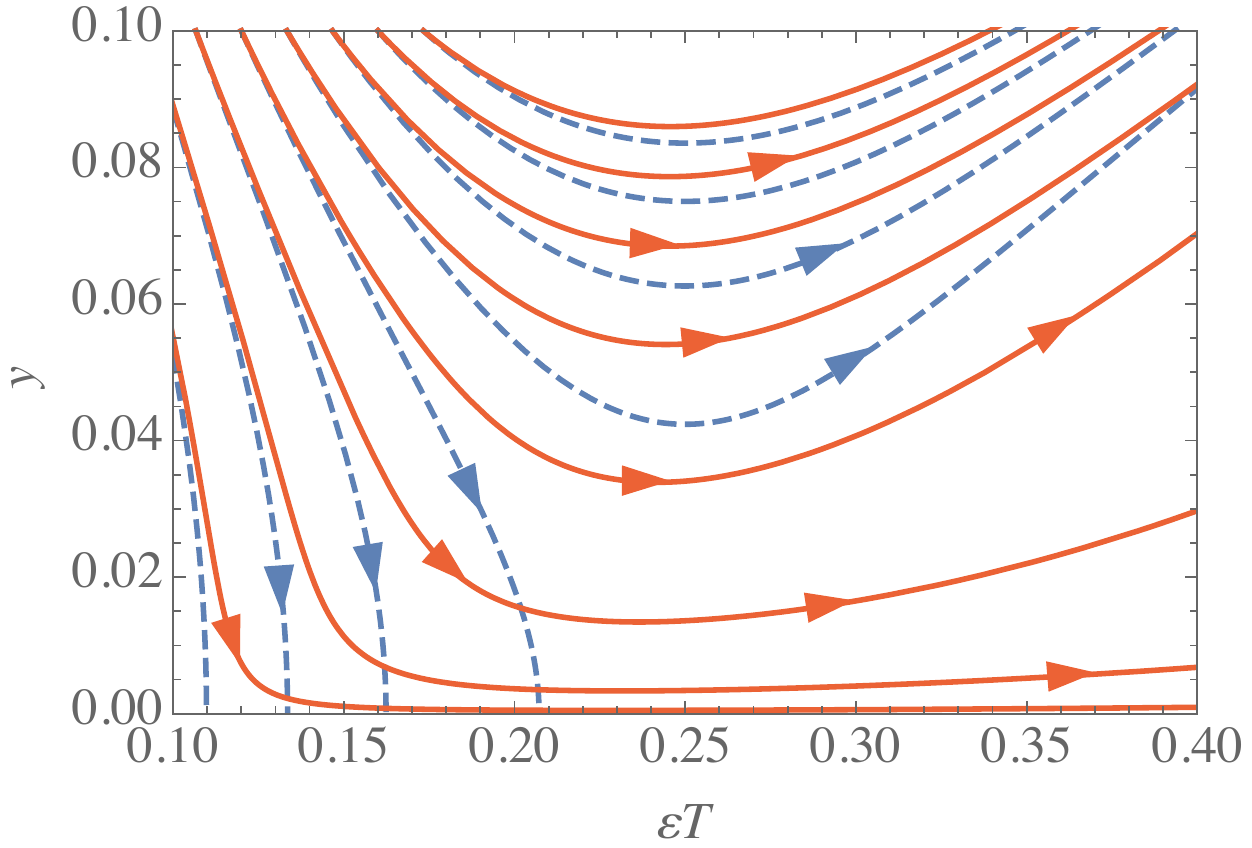} 
  \caption{(Color online.) RG flow lines as described by
    Eqs.~\eqref{eq:RGflow}. Dashed: $\lambda=0$, KT flow; Solid:
    $\lambda^2/D^2 = 0.4$.}
  \label{fig:RG}
\end{figure}

The RG flow projected onto the two parameter space of $y$ versus $\varepsilon T$
is shown in Fig.~\ref{fig:RG}. At a finite value of $\lambda$, the flow is seen
to initially closely follow the KT flow at $\lambda=0$, approaching the fixed
line at $y=0$, but ultimately departing from it in a flow toward the high
temperature plasma phase.

\begin{figure}
  \centering  
  \includegraphics[width=0.9\linewidth]{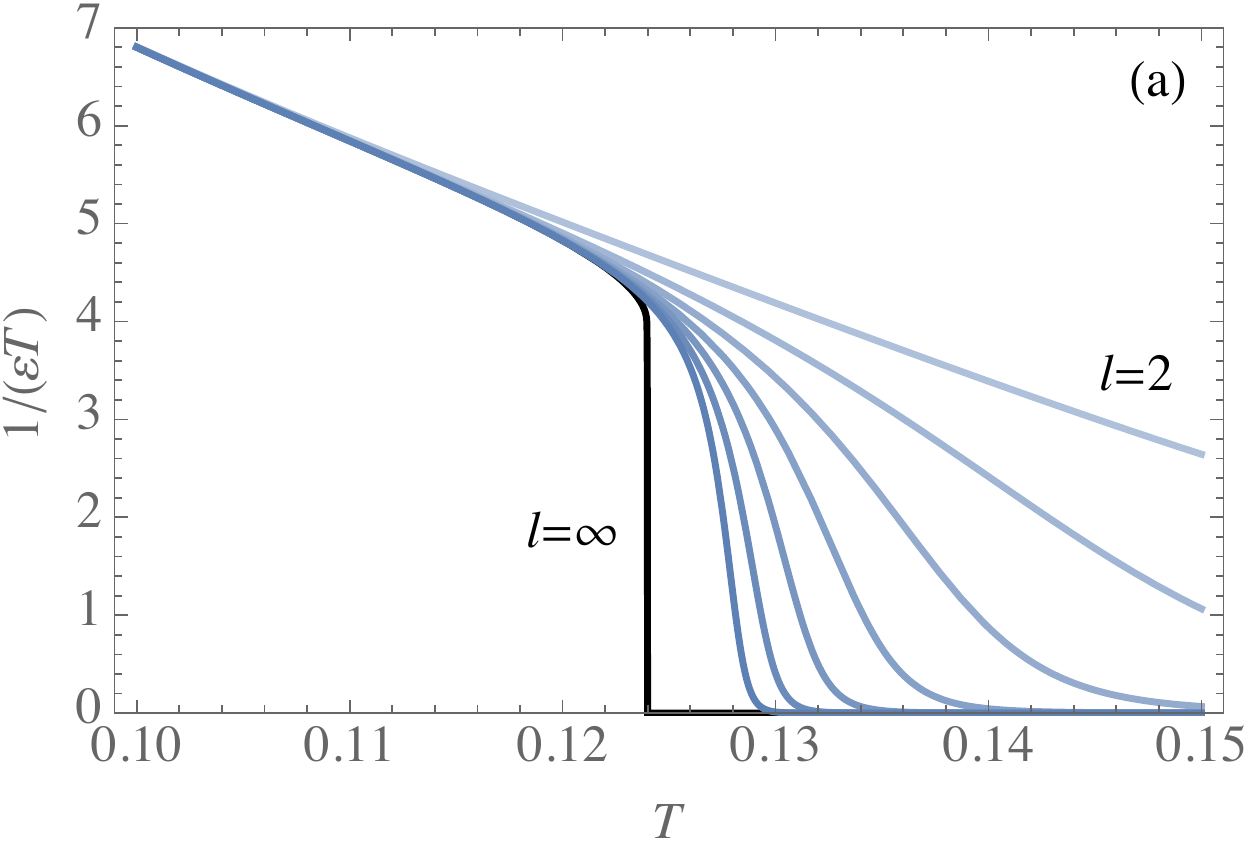}
  \includegraphics[width=0.9\linewidth]{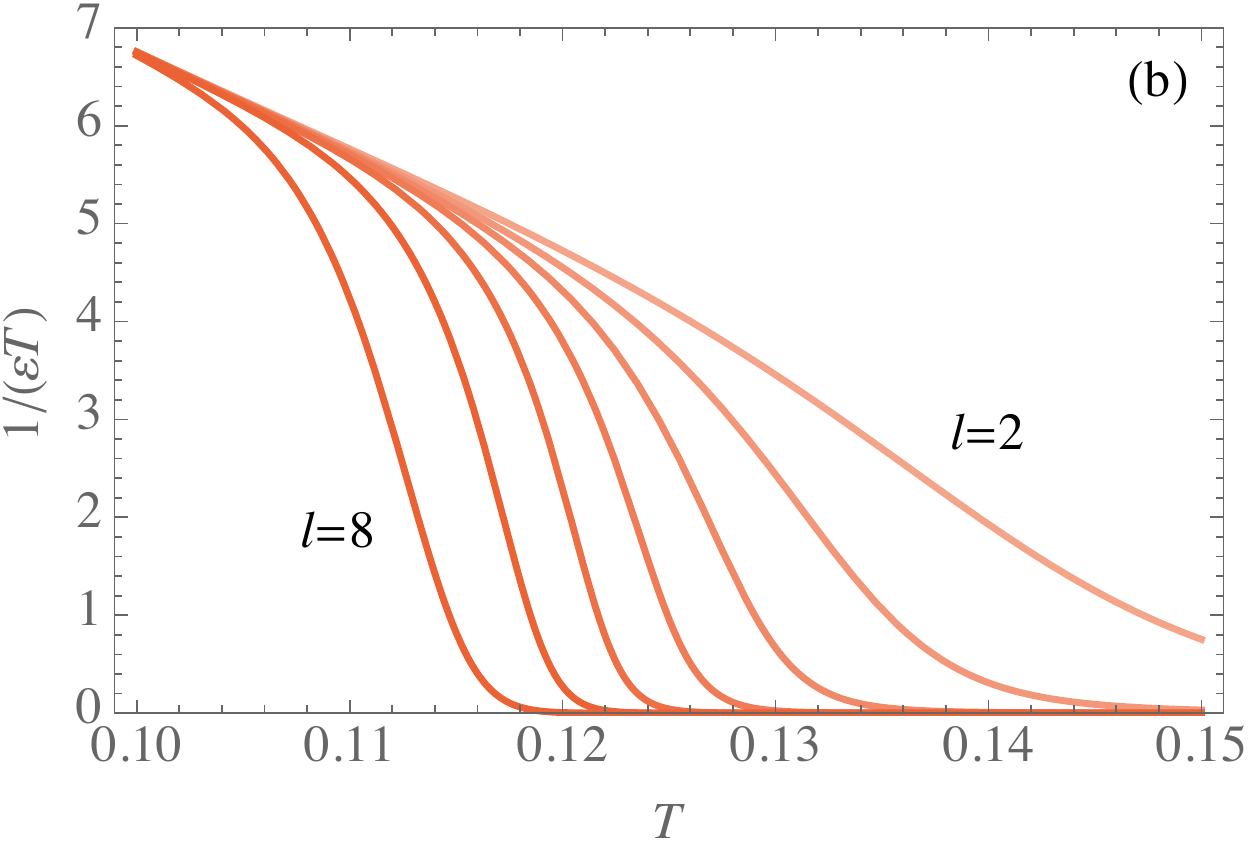}
  \caption{(Color online.) Renormalized dilectric constant in a finite-size
    system, obtained by integrating the RG flow equations~\eqref{eq:RGflow}. (a)
    2D superfluid in thermal equilibrium, i.e., with $\lambda = 0$. From top to
    bottom, the lines correspond to increasing system sizes.
    $l = 2, 3, \dotsc, 8, \infty$ denotes the logarithmic system size
    $l = \ln(L/a)$. In the thermodynamic limit, the dielectric constant
    undergoes a discontinuous jump at the critical temperature from zero to the
    universal value $1/(\varepsilon T) = 4$. (b) Non-equilibrium RG flow with
    $\lambda^2/D^2 = 0.15$. The qualitative behavior of $1/(\varepsilon T)$
    closely resembles the equilibrium one of panel (a) up to moderate system
    sizes. For small values of $\lambda/D$, the agreement is even quantitative,
    and in finite-size systems equilibrium and driven-dissipative cases are
    practically indistinguishable. The non-equilibrium character of the
    underlying dynamics becomes manifest only in the thermodynamic limit, when
    $1/(\varepsilon T)$ is renormalized to zero. In both panels, $T$ is the bare
    value of the temperature, while $\varepsilon$ is the renormalized dielectric
    constant.}
  \label{fig:sfd}
\end{figure}

Figure~\ref{fig:sfd} shows the dependence of $\varepsilon^{-1}$ (equal to the
superfluid stiffness) on the (bare) effective vortex temperature $T$ computed from the
RG flow for different system sizes. In practice, the bare effective temperature
is related to the noise appearing in the KPZ equation. It is varied by tuning
the pump power as described in Ref.~\cite{Altman2015}, where increasing pump
power reduces this bare temperature. In an infinite system at thermal
equilibrium (i.e., $\lambda=0$) the quantity $1/(\varepsilon T)$ undergoes a
universal jump from zero to $1/(\varepsilon T) = 4$ upon crossing the $KT$
transition temperature. This sharp feature is of course smeared in finite size
systems, which are seen to very slowly approach this step as their size is
increased (Fig.~\ref{fig:sfd} (a)). In the driven system ($\lambda\ne 0$),
$\epsilon^{-1}$ always vanishes in the thermodynamic limit regardless of the
bare temperature. However, the finite size scaling behavior is non-trivial. If
$\lambda$ is not too small the distinction from the $KT$ scaling behavior is
already apparent at reasonable system sizes: rather than converging to a
universal step at a critical $T$, the step is seen to be receding toward zero
temperature.

Note that in Fig.~\ref{fig:sfd} we are plotting the product of the
\emph{renormalized} dielectric constant $\varepsilon$ with the
\emph{microscopic} or bare value of the effective vortex temperature $T$. In experiments the
microscopic value of $T$ will be known as it is directly controlled by tuning
the pump power as explained above. On the other hand, the inverse dielectric
constant corresponds to the superfluid stiffness. A macroscopic measurement of
the stiffness can be implemented, as discussed above, by imposing a vortex at
the origin and measuring the decay of the physical current at a long distance $R$
from it. This will give the fully renormalized value of $\varepsilon^{-1}$.

In exciton-polariton condensates, vortices can be imprinted both under
conditions of coherent~\cite{Sanvitto2010,Boulier2016} and incoherent
pumping~\cite{Dall2014}. For a scheme to measure the current see
Ref.~\cite{Keeling2016}. From the current, the response function~\eqref{eq:4}
and hence the superfluid density can be reconstructed.

Another signature of the unbinding of vortices is the suppression of algebraic
order on scales beyond the phase boundary of the finite-size phase diagram in
Fig.~\ref{fig:phase_diagram}. On larger scales, we expect correlations to decay
rapidly due to the dephasing that is caused by vortices. In thermal equilibrium,
the decay of correlations in the high-temperature phase above the KT transition
can be shown to be exponential by means of a high-temperature
expansion~\cite{Chaikin1995} which, however, cannot be generalized
straightforwardly to the strong noise limit under non-equilibrium
conditions. Still it is reasonable to assume that the form of the decay of
correlations due to vortices is still exponential for finite values of
$\lambda$. The main challenge for observing the suppression of algebraic order
experimentally with exciton-polaritons is due to the large length scale at which
it is expected to occur --- note, however, that generically the phase boundary
in the finite size phase diagram Fig.~\ref{fig:phase_diagram} is well below
$L_*$, making the observation of the destruction of order more promising than
originally anticipated based on a treatment that neglected
vortices~\cite{Altman2015}. The precise location of the phase boundary is
controlled by the parameter lambda. As discussed in~\cite{Altman2015}, in
incoherently pumped systems this parameter is typically small, which was also
assumed in our treatment. Quite intuitively, for this parameter to become
sizable, one should use a cavity of lower quality, thus making polaritons
short-lived and hindering thermalization. We can obtain a rough estimate of the
location of the phase boundary based on the relations between the parameters in
a more microscopic model of lower polaritons coupled to a high-energy near
excitonic reservoir~\cite{Wouters2007a} and the parameters of the KPZ equation
derived in Ref.~\cite{Altman2015}. These relations lead to the following
expressions for the (absolute value of) $\lambda/D$ and the vortex temperature
$T$:
\begin{align}
  \label{eq:18}
  \frac{\lambda}{D} & = \frac{2 \bar{\gamma}}{1 + x}, \\
  \label{eq:19}
  T & = \frac{\Delta}{D} = \frac{\bar{u} \bar{\gamma}^2}{2 x \left( 1 + x \right)}
      \left[ 1 + \frac{\left( 1 + x \right)^2}{\bar{\gamma}^2} \right].
\end{align}
Here, $x$ is the dimensionless pumping strength, and $\bar{\gamma}$ and
$\bar{u}$ are dimensionless parameters which, for the experimental parameters
reported in~\cite{Lagoudakis2008}, take the values $\bar{\gamma} \approx 0.1$
and $\bar{u} \approx 0.14$~\cite{Altman2015}. The mean-field condensation
threshold is at $x = 0$ where $T$ diverges. Note that also the KPZ non-linearity
$\lambda$ depends on the pumping strength.

\begin{figure}
  \centering  
  \includegraphics[width=0.9\linewidth]{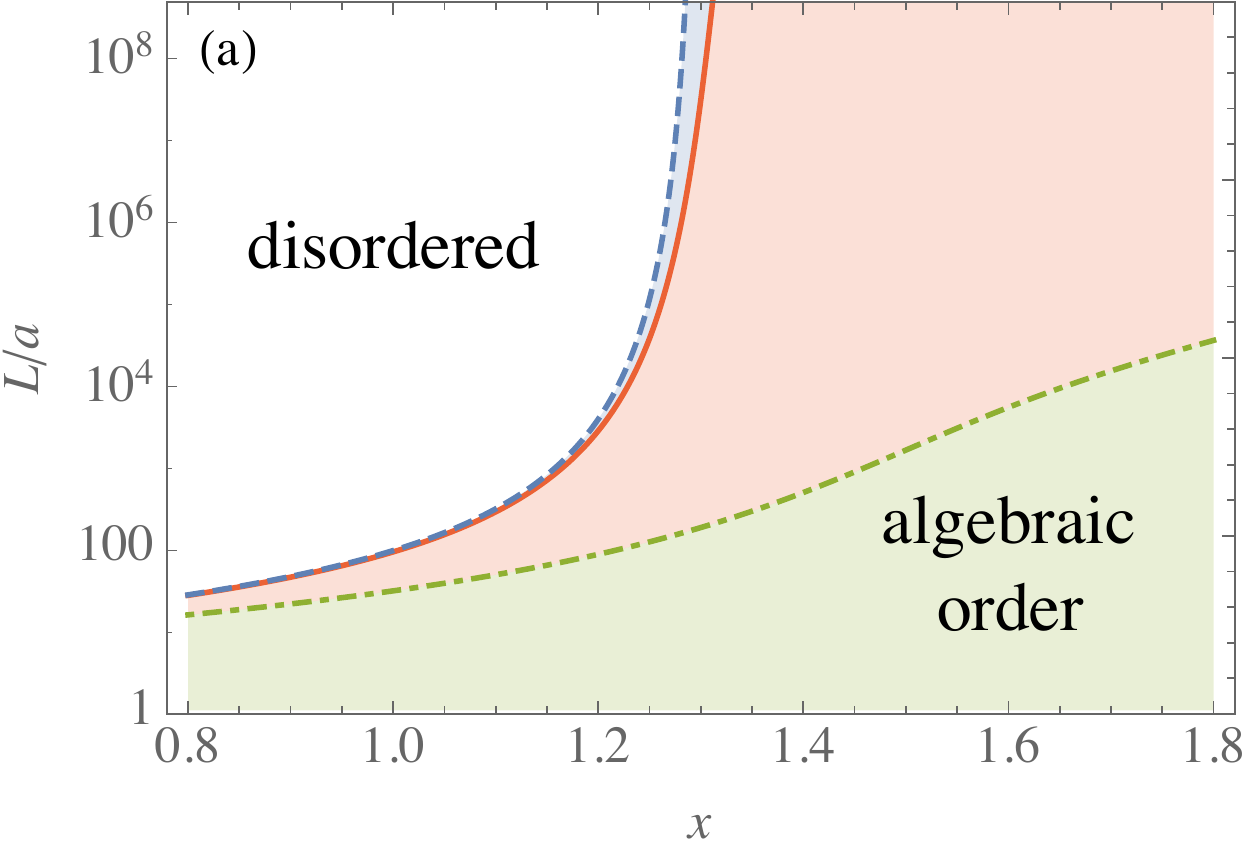}
  \includegraphics[width=0.9\linewidth]{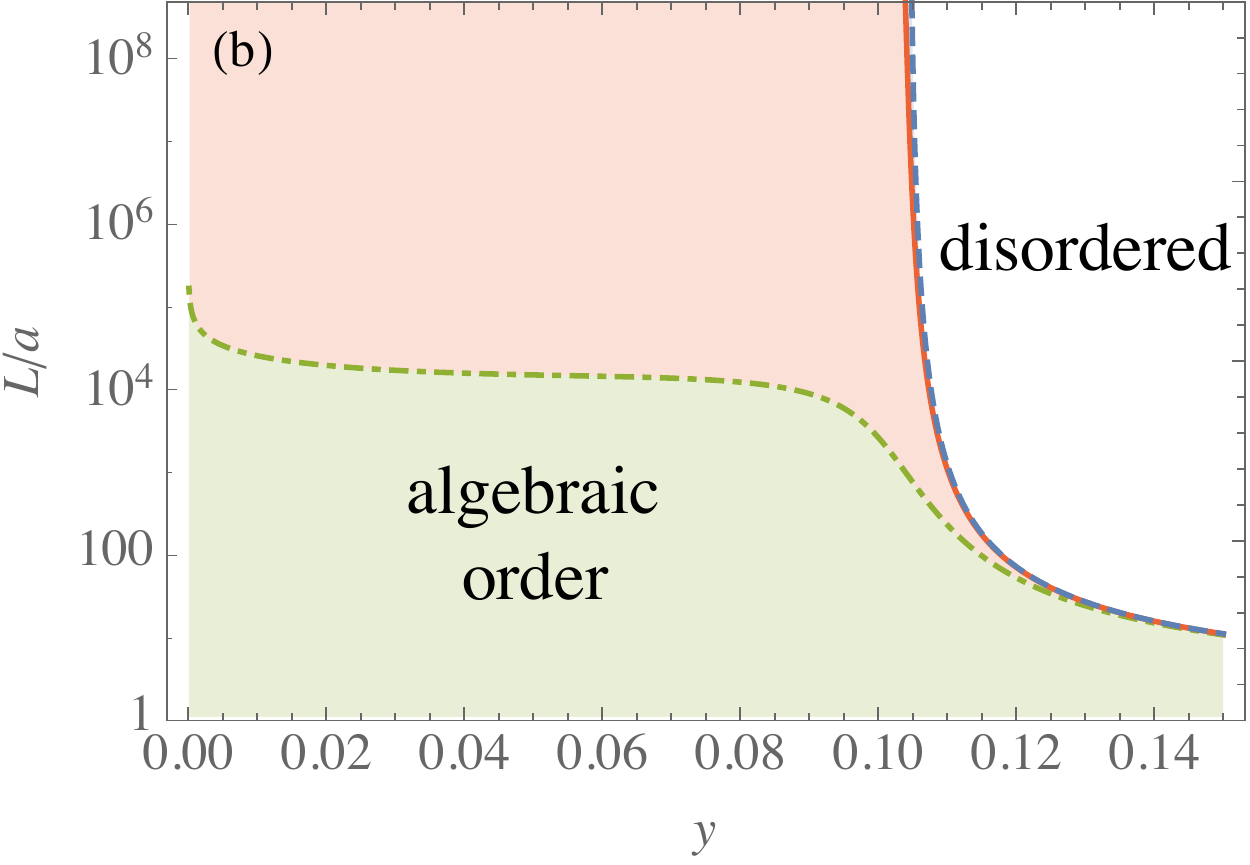}  
  \caption{(Color online.) Finite size phase diagram as a function of (a) the
    dimentionless pumping strength, $x$, and (b) the vortex fugacity $y$. In
    both panels, the red, solid line denotes the boundary between disordered and
    algebraic order phases for $\bar\gamma=0.1$ and $\bar u = 0.14$, the green,
    dotted-dashed line denote the same boundary with $\bar\gamma=0.5$, and the
    blue, dashed line is the KT phase boundary. In exciton-polaritons, the
    microscopic scale $a$ can be estimated as the healing length, which is
    typically $a \approx 1 \, \mu \mathrm{m}$, while typical system sizes are on
    the order of $L \approx 100 \, \mu \mathrm{m}$.}
  \label{fig:phase_diagram_XP}
\end{figure}

Figure~\ref{fig:phase_diagram_XP} shows the finite-size phase diagram for
incoherently pumped exciton-polaritons. The phase boundaries between
algebraically ordered (shaded regions) and disordered phases are plotted
against the pumping strength $x$ in panel (a) and the fugacity in panel (b). In
both panels, the red, solid line corresponds to the values of $\bar{\gamma}$ and
$\bar{u}$ given above. For comparison, the blue dashed line is obtained from the
equilibrium KT flow. The small values of $\lambda/D$ (for $x$ in the range from
$0.8$ to $1.8$ as shown in the figure $\lambda/D$ takes values from $0.07$ to
$0.11$) lead to a phase boundary that follows very closely the equilibrium
result. Above the equilibrium critical pump strength
$x_{\mathrm{KT}} \approx 1.30$, the phase boundary is beyond any reasonable
system size (note the logarithmic scale). A significant decrease of the phase
boundary can be achieved by increasing the value of $\lambda/D$ which, according
to Eq.~\eqref{eq:18}, necessitates a higher value of $\bar{\gamma}$. Already an
increase of $\bar{\gamma}$ by a factor of $5$ (corresponding to an increase of
the cavity loss rate by the same factor) leads to a substantial decrease of the
phase boundary (see the green, dotted-dashed line in
Fig.~\ref{fig:phase_diagram_XP}).

The estimates of the phase boundary presented in Fig.~\ref{fig:phase_diagram_XP}
(a) are obtained by integrating the RG flow equations~\eqref{eq:RGflow} from the
\textit{ad hoc} chosen microscopic value $y = 0.1$ up to $y = 1$. The dependence
of these estimates on the microscopic value of the fugacity is shown in panel
(b) of the same figure. Here, $x$ is set to the critical value of panel
(a). Note that when $y$ is below its critical value in equilibrium, the phase
boundary for the increased valued of $\bar{\gamma}$ (green, dotted-dashed line)
stays on a plateau for a wide range of values of $y$, indicating that at least
in this parameter regime our estimate of the phase boundary is not very
sensitive to the unknown microscopic value of $y$. In order to determine $y$
experimentally it would be necessary to be able to obtain the distribution of
vortices in single-shot experiments, which is at present not possible with
exciton-polaritons.

\section{Observability of KPZ scaling }

We have seen that vortices are always relevant, and ultimately unbind in a
driven condensate. Moreover, this occurs on a length scale $L_v$ that is
generally much smaller than the scale $L_*$ on which the strong coupling KPZ
physics would emerge in absence of topological defects. Hence, one might be
tempted to conclude that the physics of the strong coupling KPZ fixed point is
preempted by vortex proliferation and therefore irrelevant for two-dimensional
driven condensates.  However, our discussion so far has addressed only static
properties, such as superfluid stiffness and spatial correlations strictly at
steady state. In order to assess if KPZ scaling can be seen in experiments (or
numerics based on evolving stochastic equations for the polariton amplitude,
cf.\ e.g.~\cite{Dagvadorj2015}) measuring time-dependent quantities, we must
consider the dynamics of vortex unbinding.

As an example of a dynamical measurement we consider a quench experiment similar
to that discussed in Ref.~\cite{He2015}. The system is first pumped coherently
to prepare a condensate with an imprinted phase pattern. Hence the system is
initialized as an almost perfect condensate with no free topological defects
besides those that are imprinted externally.  A useful initial configuration, at
least for a \textit{Gedankenexperiment}, has a vortex in a hole in the middle of
the sample. From time $t=0$ onward the coherent pumping (i.e., phase imprinting)
stops, leaving only the incoherent drive. The experiment then monitors the
development of the spatial correlations as a function of time. 

We need to consider two important time scales: (i) the time $\tau_v$
it takes for vortices to unbind and proliferate (or to nucleate at the
boundary), leading to a decay of the superfluid current, should be
compared with (ii) the time scale $\tau_*$ for development of the KPZ
correlations. The characteristic time on which strong coupling KPZ
scaling emerges in a system with no topological defects is related to
the length scale $L_*$ by simple diffusive scaling
\begin{equation}
  \label{eq:13}
  \tau_*= D^{-1} L_*^2 \approx a^2 D^{-1} e^{16\pi D^3\over \Delta
    \lambda^2}\equiv a^2 D^{-1} e^{16\pi/g^2},
\end{equation}
where $a$ is the microscopic length scale and $g^2\equiv \lambda^2\Delta/ D^3$
is the dimensionless bare coupling constant of the KPZ
equation~\cite{Kardar1986}. When the time evolution begins, vortices are absent
or tightly bound in pairs on a microscopic scale. The time scale $\tau_v$ on
which vortices unbind can be estimated based on the stochastic
equation~\eqref{eq:langevin} for the relative coordinate of a vortex pair moving
in the effective potential~\eqref{eq:Veff} with no external $\bf{E}$
field. Hence, $\tau_v$ is estimated by the thermal Boltzmann factor associated
with overcoming the potential barrier to separate the pair by a distance $L_v$:
\begin{equation}
  \label{eq:14}
  \tau_v={L_v^2\over\mu y^2} \, e^{-\beta\ln(L_v/a)}\approx {a^2\over\mu y^2}
  (L_v/a)^{2+\beta}\approx{a^2\over\mu y^2} e^{{D\over \lambda}(2+\beta)},
\end{equation}
where $\beta\equiv 1/T\approx D/\Delta$. To have superfluidity at least on some
intermediate scales we should demand that $T \ll 1$.  Hence we can neglect the
$2$ with respect to $\beta$ above. The estimate for the ratio between the two
characteristic times is then
\begin{equation}
 % \label{eq:15}
  \tau_*/\tau_v\approx y^2\,{\mu\over D}
  \exp\left[{{1\over g^2}\left(16\pi-{\lambda\over D}\right)}\right].
\label{eq:ratio}
\end{equation}
Recall that $\lambda/D$ was the small parameter in the perturbative expansion
for the forces between the vortices. If $\lambda/D$ is taken to be of order one,
then $L_v$ approaches the microscopic scale $a$ (cf. Eq.~\eqref{eq:lv}). Hence in the natural regime we
discuss here $16\pi-\lambda/D>0$ and the exponential factor in
Eq.~\eqref{eq:ratio} above is large. Nonetheless, at least in principle the
ratio $\tau_*/\tau_v$ can be small if the vortex fugacity $y$ and relative
mobility $\mu/D$ are sufficiently small. In this limit we would be able to
observe strong coupling KPZ physics on intermediate time scales
$\tau_v\gg \tau\gg \tau_*$.

We note however that this is not natural within the framework of the complex GPE
since it is not possible to independently tune the bare values of $y$ and of
$\mu/D$ and they are not likely to be extremely small.  Hence in normal
situations we expect vortex proliferation to preempt the strong coupling KPZ
physics even in dynamic measurements.

\section{Conclusions \& outlook}
\label{sec:conclusions-outlook}
% Novel elctrodynamic duality. The interesting aspects are: dissipative and non-linear "photon" action. Obays gauge invariance with a modified gauge transformation. We have derived a flow equation describing the charge (vortex) unbinding, which is valid in the limit of low charge mobility and low density of charges.

We have developed a dual dynamical description of a driven-dissipative Bose
fluid in terms of the superfluid vortices.  This was achieved by extending the
well known duality between a Bose fluid and a Coulomb gas, which has been useful
in describing the KT transition~\cite{Kosterlitz1973,Jose1977,Jose1978,Savit1980} and the transport in
superfluid films near equilibrium~\cite{Ambegaokar1978,Ambegaokar1980,Cote1986}, to the case of a driven
dissipative system.
  
The electrodynamic theory obtained in the driven dissipative system is, however,
unusual. First, it describes dissipative non-relativistic photons, leading to
dynamics that is gauge invariant, but with a modified $U(1)$ gauge
transformation ${\bf A}\to {\bf A}+\nabla\chi$ and $\phi\to \phi-\chi$. The
second peculiarity is that the photon dynamics alone, even without charges, is
nonlinear. Indeed the non-linear photons constitute an equivalent dual
description of the standard non-linear KPZ equation. The coupling to charges
extends the theory to a complete long-wavelength description of the compact KPZ
equation.

% We note by contrast that the standard description of vortex dynamics in
% superconducting
% films~\cite{Ambegaokar1978,Ambegaokar1980,Cote1986,Minnhagen1987} does not
% include the dynamics of the photons. The dissipative dynamics of vortices is
% assumed, in this framework, to be governed by forces obtained from the static
% Coulomb gas duality. But, as we mentioned above, in the driven-dissipative
% system the dynamics of the photons is crucial as it reflects the underlying KPZ
% equation~\eqref{eq:KPZ} governing the condensate phase dynamics.

The dual electrodynamic theory offers a convenient framework from which to
analyze the scaling behavior of the system and the fate of the superfluid
properties at long scales. Here, we derived RG equations analogous to the
equilibrium KT flow for the vortex fugacity. For weak non-linearity and low
noise, the initial flow is similar to the equilibrium case, with the system
appearing as a superfluid up to rather long scales. Indeed,
experiments~\cite{Roumpos2012,Nitsche2014} and
numerics~\cite{Dagvadorj2015}\footnote{We note that the numerical analysis of
  Ref.~\cite{Dagvadorj2015} was performed assuming pumping in the optical
  parametric oscillator regime in which some amount of spatial anisotropy is
  imprinted by the pump wavevector.} done with limited system sizes have seen
indications of the KT physics.  However, our analysis predicts that beyond a
scale $L_v = a \exp(2 D/\lambda)$, the vortex fugacity inevitably becomes
relevant and leads to unbinding for arbitrarily low noise. In fact, within our
approximation this scale is independent of the noise level. This reflects the
fact that the present vortex unbinding mechanism relates to the deterministic
forces becoming repulsive at large distances, in contrast to the entropy driven
unbinding in a purely attractive potential in the equilibrium KT problem. In the
natural parameter regime of these systems, this vortex proliferation preempts
the establishment of correlations characterizing the strong coupling KPZ fixed
point, hence we do not expect the latter to be observable.

We note that previous work has already pointed out the screening of
the Coulomb interactions between vortices by the non-linear medium
beyond the characteristic scale $L_v = a \exp(2
D/\lambda)$~\cite{Aranson1998,Aranson1993}. This work has used a
direct asymptotic solution of the two vortex problem at large
separations. Our description in terms of an effective electrodynamic
field theory facilitates a detailed description of the universal
crossovers between the different regimes on scales below $L_v$ where
perturbation theory is valid, allowing for example, calculation of
measurable quantities such as the finite size scaling of the
superfluid stiffness (see Fig.~\ref{fig:sfd}). The same framework
allows one to compute dynamical response functions of the driven fluid
as has been done for equilibrium
superfluids~\cite{Ambegaokar1978,Ambegaokar1980,Cote1986}.

While we find that the isotropic Bose fluid inevitably loses its superfluid
properties at long scales, an intriguing question for future research is whether
a true superfluid can exist under different driving conditions. For example, we
have noted previously~\cite{Altman2015} that with a suitable anisotropy, the KPZ
non-linearity becomes irrelevant, at least when the vortex dynamics is
neglected. Such an anisotropy could potentially be realized in coherently driven
systems in the optical parametric oscillator regime. The framework developed
here should allow one to address this problem systematically, accounting both
for the anisotropic non-linear phase fluctuations together with the dynamics of
the topological defects in this anisotropic medium.

\section*{Acknowledgments}
\label{sec:acknowledgments}

We are grateful to M. Baranov, L. He and M. Szymanska for useful discussions. G.~W. and
E.~A. acknowledge support from ISF under grant number
1594-11. L.~S. and E.~A. acknowledge funding from ERC synergy grant
UQUAM, and L.~S. acknowledges support from the Koshland fellowship at
the Weizmann Institute. G.~W. was additionally supported by the NSERC
of Canada, the Canadian Institute for Advanced Research, and the
Center for Quantum Materials at the University of Toronto.
S.~D. acknowledges funding through the Institutional Strategy of the
University of Cologne within the German Excellence Initiative (ZUK 81)
and the European Research Council (ERC) under the European UnionÕs
Horizon 2020 research and innovation programme (grant agreement No
647434).

\appendix

\section{Interaction of a vortex-antivortex pair}
\label{sec:inter-vort-antiv}

In this appendix we calculate the tree-level corrections to the interaction
between a vortex ($n_+ = 1$) at $\mathbf{r}_+$ and an antivortex ($n_- = - 1$)
at $\mathbf{r}_-$, given by Eqs.~\eqref{eq:TL1} and~\eqref{eq:TL2}. To zeroth
order in $\lambda$, the force between between a vortex and an antivortex at a
distance $R = \abs{\mathbf{R}}$, where $\mathbf{R} = \mathbf{r}_+ -
\mathbf{r}_-$, which is larger than the short-distance cutoff $a$, is given by
\begin{equation}
  \label{eq:tralala32}
  \mathbf{f}_0(\mathbf{R}) = \frac{2 \pi}{\varepsilon} \nabla G(\mathbf{R}) = -
  \frac{\mathbf{R}}{\varepsilon R^2},
\end{equation}
where $G(\mathbf{R})$ is (minus) the inverse of the Laplacian in 2D, see
Eq.~\eqref{eq:G}. At distances shorter than the cutoff $a$ we set the force to
zero. The corrections to the zeroth order force $\mathbf{f}_0(\mathbf{R})$ are
given in terms of integrals which are in part divergent in the limit
$a/R \to 0$. The long-distance behavior of the interaction is determined by the
leading terms in an asymptotic expansion of the integrals for $a/R \to 0$ which
we calculate below.

\subsection{First order correction}
\label{sec:first-order-corr}

For future convenience, we write the first order correction as
$\mathbf{f}^{(1)}(\mathbf{r}_+ - \mathbf{r}_-) = -\mathbf{g}(\mathbf{r}_+)$. The
quantity $\mathbf{g}(\mathbf{r})$ appears also in the calculation of the second
order correction in Sec.~\ref{sec:second-order-corr} in below, and as can be
seen in Eq.~\eqref{eq:TL1} it is given by
\begin{equation}
  \label{eq:tralalavortex-int-3}
  \begin{split}
    \mathbf{g}(\mathbf{r}) & = \frac{\varepsilon \lambda}{4 \pi D}
    \int_{\mathbf{r}'} \unitvec{z} \times \mathbf{f}_0(\mathbf{r}' - \mathbf{r})
    \abs{\mathbf{f}_0(\mathbf{r}' - \mathbf{r}_+) - \mathbf{f}_0(\mathbf{r}' -
      \mathbf{r}_-)}^2 \\ & = \frac{\lambda}{2 D \varepsilon^2} \unitvec{z}
    \times \mathbf{a}(\mathbf{r}).
  \end{split}
\end{equation}
In the last equality, we introduced another auxiliary quantity
$\mathbf{a}(\mathbf{r})$, which can be decomposed as
\begin{equation}
  \label{eq:tralalavortex-int-4}  
  \mathbf{a}(\mathbf{r}) = \mathbf{a}_+(\mathbf{r}) - 2
  \mathbf{a}_{+-}(\mathbf{r}) + \mathbf{a}_-(\mathbf{r}),
\end{equation}
where
\begin{align}
  \label{eq:tralala3}
  \mathbf{a}_{\pm}(\mathbf{r})
  & = \frac{\varepsilon^3}{2 \pi}
    \int_{\mathbf{r}'} \mathbf{f}_0(\mathbf{r}' - \mathbf{r})
    \abs{\mathbf{f}_0(\mathbf{r}' - \mathbf{r}_{\pm})}^2, \\
  \label{eq:tralala34}
  \mathbf{a}_{+-}(\mathbf{r})
  & = \frac{\varepsilon^3}{2 \pi}
    \int_{\mathbf{r}'} \mathbf{f}_0(\mathbf{r}' - \mathbf{r}) \left(
    \mathbf{f}_0(\mathbf{r}' - \mathbf{r}_+) \cdot \mathbf{f}_0(\mathbf{r}' -
    \mathbf{r}_-) \right).
\end{align}
Instead of calculating $\mathbf{a}(\mathbf{r})$ for arbitrary values of
$\mathbf{r}$, we first focus on the relevant case for the first order
correction, i.e., $\mathbf{r} = \mathbf{r}_+$. Let's start with the contribution
$\mathbf{a}_+(\mathbf{r}_+)$ defined in Eq.~\eqref{eq:tralala3}. To make
progress with this integral, we shift the integration variable
$\mathbf{r}' \to \mathbf{r}' + \mathbf{r}_+$ and drop the dash, i.e., relabel
$\mathbf{r}' \to \mathbf{r}.$ After these manipulations, it can readily be seen
that $\mathbf{a}_+(\mathbf{r}_+)$ vanishes due to the rotational symmetry of the
integrand,
\begin{equation}
  \label{eq:tralala4}  
  \mathbf{a}_+(\mathbf{r}_+) = \frac{\varepsilon^3}{2 \pi} \int_{\mathbf{r}}
  \mathbf{f}_0(\mathbf{r}) \abs{\mathbf{f}_0(\mathbf{r})}^2 =
  - \frac{1}{2 \pi} \int_{\mathbf{r}} \frac{\mathbf{r}}{r^4} = 0.
\end{equation}
(Note that the divergence at $\mathbf{r} = 0$ is regularized by the
short-distance cutoff $a$.) Next we consider $\mathbf{a}_-(\mathbf{r}_+)$, which
requires us to do some actual work. Shifting
$\mathbf{r}' \to \mathbf{r}' + \mathbf{r}_-$ and renaming
$\mathbf{r}' \to \mathbf{r}$ as above, and moreover denoting the relative
coordinate as $\mathbf{R} = \mathbf{r}_+ - \mathbf{r}_-$, we have
\begin{equation}
  \label{eq:tralala5}
  \mathbf{a}_-(\mathbf{r}_+) = \frac{\varepsilon^3}{2 \pi} \int_{\mathbf{r}}
  \mathbf{f}_0(\mathbf{r} - \mathbf{R}) \abs{\mathbf{f}_0(\mathbf{r})}^2 = -
  \frac{1}{2 \pi} \int_{\mathbf{r}} \frac{\mathbf{r} -
    \mathbf{R}}{\abs{\mathbf{r} - \mathbf{R}}^2} \frac{1}{r^2}.
\end{equation}
As we will show minutely below, the pole at $\mathbf{r} = 0$ gives a logarithmic
contribution, while the apparent divergence at $\mathbf{r} = \mathbf{R}$ is
lifted by the angular integration as in Eq.~\eqref{eq:tralala4}. This can be
seen by simplifying the integrand in the vicinity of the pole, i.e., by
replacing $1/r^2 \to 1/R^2$, and shifting the integration variable back to
$\mathbf{r} \to \mathbf{r} + \mathbf{R}$, thus moving the pole to
$\mathbf{r} = 0$. Then, as above,
$\int_{\mathbf{r}} \left( \mathbf{r}/r^2 \right) = 0$. Hence there is no need to
keep a finite short-distance cutoff $a$ at this pole if we agree to perform the
angular integration before the radial one. In order to carry out the integral
explicitly, we parameterize $\mathbf{r}$ and $\mathbf{R}$ in polar coordinates
as
\begin{equation}
  \label{eq:tralalavortex-int-17}
  \mathbf{r} = r
  \begin{pmatrix}
    \cos(\theta + \theta_{\mathbf{R}}) \\ \sin(\theta + \theta_{\mathbf{R}})
  \end{pmatrix},
  \quad \mathbf{R} = R
  \begin{pmatrix}
    \cos(\theta_{\mathbf{R}}) \\ \sin(\theta_{\mathbf{R}})
  \end{pmatrix}.
\end{equation}
In this representation, the scalar product between $\mathbf{r}$ and $\mathbf{R}$
is just $\mathbf{r} \cdot \mathbf{R} = r R \cos(\theta)$, and the integral takes
the form
\begin{multline}
  \label{eq:tralala6}  
  \mathbf{a}_-(\mathbf{r}_+) = - \frac{1}{2 \pi} \int_a^{\infty} \frac{dr}{r}
  \int_0^{2 \pi} d \theta \frac{1}{r^2 + R^2 - 2 r R \cos(\theta)} \\ \times
  \left( r
    \begin{pmatrix}
      \cos(\theta) \cos(\theta_{\mathbf{R}}) - \sin(\theta)
      \sin(\theta_{\mathbf{R}}) \\
      \sin(\theta) \cos(\theta_{\mathbf{R}}) + \cos(\theta)
      \sin(\theta_{\mathbf{R}})
    \end{pmatrix}
    - \mathbf{R} \right).  
\end{multline}
Due to their rotational symmetry, the terms involving $\sin(\theta)$ vanish. The
remaining integrals can easily be performed with the aid of
Ref.~\cite{Gradshteyn2007}, leading to, for $R > a$,
\begin{equation}
  \label{eq:tralala9}
  \mathbf{a}_-(\mathbf{r}_+) = \frac{\mathbf{R}}{R^2} \ln(R/a),
\end{equation}
and $\mathbf{a}_-(\mathbf{r}_+) = 0$ for $R < a$. We move on to calculate
$\mathbf{a}_{+-}(\mathbf{r}_+)$, which is defined in
Eq.~\eqref{eq:tralala34}. Here, the by now familiar shift of integration
variables leads us to
\begin{equation}
  \label{eq:tralala10}  
  \begin{split}
    \mathbf{a}_{+-}(\mathbf{r}_+) & = \frac{\varepsilon^3}{2 \pi}
    \int_{\mathbf{r}} \mathbf{f}_0(\mathbf{r}) \left( \mathbf{f}_0(\mathbf{r})
      \cdot \mathbf{f}_0(\mathbf{r} + \mathbf{R}) \right) \\ & = - \frac{1}{2
      \pi} \int_{\mathbf{r}} \frac{\mathbf{r}}{r^4} \frac{\mathbf{r} \cdot
      \left( \mathbf{r} + \mathbf{R} \right)}{\abs{\mathbf{r} + \mathbf{R}}^2},
  \end{split}
\end{equation}
and using the same polar coordinate
representation~\eqref{eq:tralalavortex-int-17} as above we find
\begin{equation}
  \label{eq:6}
  \mathbf{a}_{+-}(\mathbf{r}_+) = - \frac{\mathbf{R}}{2 R^2} \left( \ln(R/a) -
    \frac{1}{2} \right), % \theta(R - a)  + \frac{\mathbf{R}}{4 a^2} \theta(a -
    % R).
\end{equation}
where we again assume $R > a$. Adding the contributions from
Eqs.~\eqref{eq:tralala4},~\eqref{eq:tralala9}, and~\eqref{eq:tralala10}, we find
\begin{equation}
  \label{eq:tralala11}
  \mathbf{a}(\mathbf{r}_+) = \frac{\mathbf{R}}{R^2} \left( 2 \ln(R/a) -
    \frac{1}{2} \right),
\end{equation}
which upon insertion in Eq.~\eqref{eq:tralalavortex-int-3} yields the first
order correction to the interaction in Eq.~\eqref{eq:1}.

\subsection{Second order correction}
\label{sec:second-order-corr}

The second order correction to the force, given by Eq.~\eqref{eq:TL2}, can be
written as
\begin{multline}
  \label{eq:tralala26}  
  \mathbf{f}^{(2)}(\mathbf{R}) = - \frac{\varepsilon \lambda}{4 \pi D}
  \int_{\mathbf{r}} \unitvec{z} \times \mathbf{f}_0(\mathbf{r}) \\ \times \left[
    \left( \mathbf{f}_0(\mathbf{r}) - \mathbf{f}_0(\mathbf{r} + \mathbf{R})
    \right) \cdot \mathbf{g}(\mathbf{r} + \mathbf{r}_+) \right],
\end{multline}
which shows that we are now required to evaluate $\mathbf{g}(\mathbf{r})$ (and
hence $\mathbf{a}(\mathbf{r})$) for arbitrary values of $\mathbf{r}$ (in
particular, for values different from $\mathbf{r} = \mathbf{r}_+$). To
facilitate the rather tedious calculation of the second order correction, we
break the latter up into several contributions. To wit, we decompose
$\mathbf{g}(\mathbf{r})$ in Eq.~\eqref{eq:tralala26} according to
$\mathbf{g}(\mathbf{r}) = \mathbf{g}_1(\mathbf{r}) + \mathbf{g}_2(\mathbf{r})$,
where
\begin{align}
  \label{eq:tralala35}
  \mathbf{g}_1(\mathbf{r})
  & = \frac{\lambda}{2 D \varepsilon^2} \unitvec{z}
    \times \left( \mathbf{a}_+(\mathbf{r}) + \mathbf{a}_-(\mathbf{r}) \right), \\  
  \label{eq:tralala28}
  \mathbf{g}_2(\mathbf{r})
  & = - \frac{\lambda}{D \varepsilon^2}
    \unitvec{z} \times \mathbf{a}_{+-}(\mathbf{r}).
\end{align}
$\mathbf{a}_{\pm}(\mathbf{r})$ and $\mathbf{a}_{+-}(\mathbf{r})$ are defined in
Eqs.~\eqref{eq:tralala3} and~\eqref{eq:tralala34}, respectively, and the are
related to $\mathbf{g}(\mathbf{r})$ by Eq.~\eqref{eq:tralalavortex-int-3}. The
decomposition of $\mathbf{g}(\mathbf{r})$ entails a corresponding one of the
second order correction as
$\mathbf{f}^{(2)}(\mathbf{R}) = \mathbf{f}^{(2)}_1(\mathbf{R}) +
\mathbf{f}^{(2)}_2(\mathbf{R})$.

First we consider the first term, i.e.,
$\mathbf{f}^{(2)}_1(\mathbf{R})$. According to Eq.~\eqref{eq:tralala35}, we have
to calculate $\mathbf{a}_{\pm}(\mathbf{r})$. This calculation proceeds along the
lines of the one of $\mathbf{a}_-(\mathbf{r}_+)$ presented above, resulting in
\begin{equation}
  \label{eq:tralala13}
  \mathbf{a}_{\pm}(\mathbf{r}) = \frac{\mathbf{R}_{\pm}}{R^2_{\pm}}
  \ln(R_{\pm}/a) = - \varepsilon \mathbf{f}_0(\mathbf{R}_{\pm}) \ln(R_{\pm}/a),
\end{equation}
where we defined $\mathbf{R}_{\pm} = \mathbf{r} - \mathbf{r}_{\pm}$. As in
Eq.~\eqref{eq:tralala9} this expression is cut off at distances $R_{\pm} < a$.
Inserting Eq.~\eqref{eq:tralala13} in Eq.~\eqref{eq:tralala35}, and the latter
in Eq.~\eqref{eq:tralala26}, we obtain
\begin{widetext}
  \begin{equation}
  \label{eq:tralala15}
  \begin{split}
    \mathbf{f}^{(2)}_1(\mathbf{R}) & = \frac{\lambda^2}{8 \pi D^2}
    \int_{\mathbf{r}} \unitvec{z} \times \mathbf{f}_0(\mathbf{r}) \left\{ \left(
        \mathbf{f}_0(\mathbf{r}) - \mathbf{f}_0(\mathbf{r} + \mathbf{R}) \right)
      \cdot \left[ \unitvec{z} \times \left( \mathbf{f}_0(\mathbf{r}) \ln(r/a) +
          \mathbf{f}_0(\mathbf{r} + \mathbf{R}) \ln(\abs{\mathbf{r} +
            \mathbf{R}}/a) \right) \right] \right\} \\ & = - \frac{1}{2 \pi}
    \left( \frac{\lambda}{2 D} \right)^2 \int_{\mathbf{r}} \ln(r \abs{\mathbf{r}
      + \mathbf{R}}/a^2) \unitvec{z} \times \mathbf{f}_0(\mathbf{r}) \left[
      \unitvec{z} \cdot \left( \mathbf{f}_0(\mathbf{r}) \times
        \mathbf{f}_0(\mathbf{r} + \mathbf{R}) \right) \right].
  \end{split}
\end{equation}
\end{widetext}
In order to single out the parts of this integral which diverge for $a/R \to 0$,
we rewrite the logarithm in the integrand as the sum of two terms,
\begin{equation}
  \label{eq:tralala20}  
  \ln(r \abs{\mathbf{r} + \mathbf{R}}/a^2) = \ln(r R/a^2) +
  \ln(\abs{\mathbf{r} + \mathbf{R}}/R),  
\end{equation}
and introduce some more auxiliary quantities,
\begin{equation}
  \label{eq:tralala19}  
  \mathbf{b}_1(\mathbf{R})
   = \frac{\varepsilon^3}{2 \pi} \int_{\mathbf{r}}
    \ln(r R/a^2) \mathbf{f}_0(\mathbf{r}) \left[ \unitvec{z} \cdot \left(
    \mathbf{f}_0(\mathbf{r}) \times \mathbf{f}_0(\mathbf{r} + \mathbf{R})
  \right) \right],
\end{equation}
and
\begin{multline}
\label{eq:tralala42}
\mathbf{b}_2(\mathbf{R}) = \frac{\varepsilon^3}{2 \pi} \int_{\mathbf{r}}
\ln(\abs{\mathbf{r} + \mathbf{R}}/R) \\ \times \mathbf{f}_0(\mathbf{r}) \left[
  \unitvec{z} \cdot \left( \mathbf{f}_0(\mathbf{r}) \times
    \mathbf{f}_0(\mathbf{r} + \mathbf{R}) \right) \right],
\end{multline}
such that
\begin{equation}
  \label{eq:7}
  \mathbf{f}^{(2)}_1(\mathbf{R}) = - \left(
    \frac{\lambda}{2 D} \right)^2 \frac{1}{\varepsilon^3} \unitvec{z} \times
  \left( \mathbf{b}_1(\mathbf{R}) + \mathbf{b}_2(\mathbf{R}) \right).
\end{equation}
As always, we start with $\mathbf{b}_1(\mathbf{R})$. Using the polar
representation introduced in Eq.~\eqref{eq:tralalavortex-int-17} we find
$\unitvec{z} \cdot \left( \mathbf{r} \times \mathbf{R} \right) = - r R
\sin(\theta)$, which we use in Eq.~\eqref{eq:tralala19} to rewrite the latter as
\begin{widetext}
  \begin{equation}
  \label{eq:tralala21}
  \begin{split}
    \mathbf{b}_1(\mathbf{R}) & = - \frac{1}{2 \pi} \int_{\mathbf{r}} \ln(r
    R/a^2) \frac{\mathbf{r}}{r^4} \frac{\unitvec{z} \cdot \left( \mathbf{r}
        \times \mathbf{R} \right)}{\abs{\mathbf{r} + \mathbf{R}}^2} \\ & = 
    \frac{R}{2 \pi} \int_a^{\infty} \frac{d r}{r} \ln(r R/a^2) \int_0^{2 \pi} d
    \theta \frac{\sin(\theta)}{r^2 + R^2 + 2 r R \cos(\theta)}
    \begin{pmatrix}
      \cos(\theta) \cos(\theta_{\mathbf{R}}) - \sin(\theta)
      \sin(\theta_{\mathbf{R}}) \\
      \sin(\theta) \cos(\theta_{\mathbf{R}}) + \cos(\theta)
      \sin(\theta_{\mathbf{R}})
    \end{pmatrix}.
  \end{split}
\end{equation}
\end{widetext}
Due to the sine function $\sin(\theta)$ in the numerator, in the polar
representation of $\mathbf{r}$, only the terms involving a sine as well have to
be kept. The others just have the wrong behavior under $\theta \to -
\theta$. Using
\begin{equation}
  \label{eq:tralala22}
  \begin{pmatrix}
    - \sin(\theta_{\mathbf{R}}) \\ \cos(\theta_{\mathbf{R}})
  \end{pmatrix}
  = \unitvec{z} \times \unitvec{R},
\end{equation}
and Ref.~\cite{Gradshteyn2007} we find
\begin{equation}
  \label{eq:tralala23}  
  \mathbf{b}_1(\mathbf{R}) = 
  \frac{\unitvec{z} \times \mathbf{R}}{8 R^2} \left( 6 \ln(R/a)^2 + 4 \ln(R/a)
    + 1 \right).
\end{equation}
The logarithms are due to the pole of the integrand at $\mathbf{r} = 0$, whereas
the apparent pole at $\mathbf{r} = \mathbf{R}$ is lifted after performing the
angular integration as we have already seen several times above.

Our decomposition of the logarithm in Eq.~\eqref{eq:tralala20} ensures that
there are no additional logarithmic contributions in
$\mathbf{b}_2(\mathbf{R})$. This can be seen by noting that the expansion of the
logarithm in the integrand in Eq.~\eqref{eq:tralala42} for $\mathbf{r} \to 0$
yields an additional factor of $r$, and that the integrand is thus regular at
$\mathbf{r} = 0$. To calculate $\mathbf{b}_2(\mathbf{R})$ explicitly, we first
shift the integration variable according to
$\mathbf{r} \to \mathbf{r} - \mathbf{R}$ and then switch to polar coordinates as
in Eq.~\eqref{eq:tralalavortex-int-17}. The resulting integrals are similar to
the ones we have already become acquainted with, and a straightforward
evaluation yields $\mathbf{b}_2(\mathbf{R}) = 0$. Inserting this result and
Eq.~\eqref{eq:tralala23} in Eq.~\eqref{eq:7} we obtain
\begin{equation}
  \label{eq:tralala25}
  \mathbf{f}^{(2)}_1(\mathbf{R}) = \frac{1}{8} \left( \frac{\lambda}{2 D} \right)^2
  \frac{1}{\varepsilon^3} \frac{\mathbf{R}}{R^2} \left( 6 \ln(R/a)^2 + 4 \ln(R/a)
    + 1 \right).
\end{equation}

It remains to calculate $\mathbf{f}^{(2)}_2(\mathbf{R})$, which is obtained by
inserting Eq.~\eqref{eq:tralala34} in Eq.~\eqref{eq:tralala28} and the latter in
Eq.~\eqref{eq:tralala26},
\begin{multline}
  \label{eq:tralala30}  
  \mathbf{f}^{(2)}_2(\mathbf{R}) = \left( \frac{\lambda}{2 D} \right)^2
  \frac{\varepsilon^2}{2 \pi^2} \int_{\mathbf{r}, \mathbf{r}'} \unitvec{z}
  \times \mathbf{f}_0(\mathbf{r}) \\ \times \left[ \left(
      \mathbf{f}_0(\mathbf{r}) - \mathbf{f}_0(\mathbf{r} + \mathbf{R}) \right)
    \cdot \left( \unitvec{z} \times \mathbf{f}_0(\mathbf{r}' - \mathbf{r})
    \right) \right] \\ \times \left( \mathbf{f}_0(\mathbf{r}') \cdot
    \mathbf{f}_0(\mathbf{r}' + \mathbf{R}) \right).
\end{multline}
The reader won't be surprised to see that we also break up this part into two
contributions,
$\mathbf{f}^{(2)}_2(\mathbf{R})= \mathbf{f}^{(2)}_{2,1}(\mathbf{R}) +
\mathbf{f}^{(2)}_{2,2}(\mathbf{R})$, where
\begin{widetext}
  \begin{equation}
  \label{eq:tralala36}
  \begin{split}
    \mathbf{f}^{(2)}_{2,1}(\mathbf{R}) & = - \left( \frac{\lambda}{2 D} \right)^2
    \frac{\varepsilon^2}{2 \pi^2} \int_{\mathbf{r}, \mathbf{r}'} \unitvec{z}
    \times \mathbf{f}_0(\mathbf{r}) \left[ \mathbf{f}_0(\mathbf{r}) \cdot \left(
        \unitvec{z} \times \mathbf{f}_0(\mathbf{r} - \mathbf{r}') \right)
    \right] \left( \mathbf{f}_0(\mathbf{r}') \cdot
      \mathbf{f}_0(\mathbf{r}' + \mathbf{R}) \right) \\
    & = \left( \frac{\lambda}{2 D} \right)^2 \frac{1}{\pi \varepsilon}
    \int_{\mathbf{r}'} \unitvec{z} \times \mathbf{c}(\mathbf{r}') \left(
      \mathbf{f}_0(\mathbf{r}') \cdot \mathbf{f}_0(\mathbf{r}' + \mathbf{R})
    \right),
  \end{split}
\end{equation}
with
\begin{equation}
  \label{eq:tralala44}
  \mathbf{c}(\mathbf{r}') = \frac{\varepsilon^3}{2 \pi} \int_{\mathbf{r}}
  \mathbf{f}_0(\mathbf{r}) \left[ \unitvec{z} \cdot \left(
      \mathbf{f}_0(\mathbf{r}) \times \mathbf{f}_0(\mathbf{r} - \mathbf{r}')
    \right) \right],
\end{equation}
and
\begin{equation}
  \label{eq:tralala43}
  \begin{split}
    \mathbf{f}^{(2)}_{2,2}(\mathbf{R}) & = \left( \frac{\lambda}{2 D} \right)^2
    \frac{\varepsilon^2}{2 \pi^2} \int_{\mathbf{r}, \mathbf{r}'} \unitvec{z}
    \times \mathbf{f}_0(\mathbf{r}) \left[ \mathbf{f}_0(\mathbf{r} + \mathbf{R})
      \cdot \left( \unitvec{z} \times \mathbf{f}_0(\mathbf{r} - \mathbf{r}')
      \right) \right] \left( \mathbf{f}_0(\mathbf{r}') \cdot
      \mathbf{f}_0(\mathbf{r}' + \mathbf{R}) \right) \\ & = \left(
      \frac{\lambda}{2 D} \right)^2 \frac{1}{\varepsilon^3} \unitvec{z} \times
    \mathbf{d}(\mathbf{R}),
\end{split}
\end{equation}
where
\begin{equation}
  \label{eq:tralala33}
  \mathbf{d}(\mathbf{R}) = \frac{\varepsilon^5}{2 \pi^2} \int_{\mathbf{r}, \mathbf{r}'}
  \mathbf{f}_0(\mathbf{r}) \left[
    \unitvec{z} \cdot \left( \mathbf{f}_0(\mathbf{r} - \mathbf{r}') \times
      \mathbf{f}_0(\mathbf{r} + \mathbf{R}) \right) \right] \left(
    \mathbf{f}_0(\mathbf{r}') \cdot
    \mathbf{f}_0(\mathbf{r}' + \mathbf{R}) \right).
\end{equation}
\end{widetext}
The integral in Eq.~\eqref{eq:tralala44} can be evaluated using the same
techniques as before. We find the result
\begin{equation}
  \label{eq:tralala39}
  \mathbf{c}(\mathbf{r}') = - \frac{\unitvec{z}
    \times \mathbf{r}'}{4 r^{\prime 2}} \left( 2 \ln(r'/a) + 1 \right).
\end{equation}
Plugging this into Eq.~\eqref{eq:tralala36} and omitting details of the further
evaluation which is similar to the ones presented above, we almost immediately
obtain
\begin{equation}
  \label{eq:tralala40}  
  \mathbf{f}^{(2)}_{2,1}(\mathbf{R}) = 
  \frac{1}{4} \left( \frac{\lambda}{2 D} \right)^2 \frac{1}{\varepsilon^3}
  \frac{\mathbf{R}}{R^2} \left( \ln(R/a)^2 - 1 \right).
\end{equation}

Finally, we turn our attention to the single missing piece,
$\mathbf{d}(\mathbf{R})$ defined in Eq.~\eqref{eq:tralala33}. This integral
turns out to be convergent for $a/R \to 0$ and equal to zero in this limit. If
one is clever enough, one might be able to see this by carefully considering
symmetries of the integrand. We did the full calculation instead.

This undertaking is greatly facilitated by using the following Fourier-cosine
series:\footnote{We warmly thank M. Baranov for this hint.}
\begin{equation}
  \label{eq:tralala54}
  \frac{1}{\abs{\mathbf{r} - \mathbf{r}'}^2} = \frac{1}{\abs{r^2 - r^{\prime
        2}}} \sum_{n = 0}^{\infty} \left( 2 - \delta_{n,0} \right) \left(
    \frac{r_{<}}{r_{>}} \right)^n \cos(n (\theta - \theta')),
\end{equation}
where $r_{<}$ and $r_{>}$ are the lesser and greater, respectively, of $r$ and
$r'$. The angles $\theta$ and $\theta'$ in the polar coordinate representation
of $\mathbf{r}$ and $\mathbf{r}'$ are measured with respect to
$\theta_{\mathbf{R}}$ as in Eq.~\eqref{eq:tralalavortex-int-17}. Inserting the
above expansion in Eq.~\eqref{eq:tralala33} yields
\begin{widetext}
  \begin{multline}
  \label{eq:tralala55}  
  \mathbf{d}(\mathbf{R}) = - \frac{1}{2 \pi^2} \sum_{n = 0}^{\infty} \left( 2 -
    \delta_{n, 0} \right) \int_0^{\infty} dr \frac{1}{r^2 + R^2} \int_0^{\infty}
  dr' \frac{1}{r^{\prime 2} + R^2} \frac{1}{\abs{r^2 - r^{\prime 2}}} \left(
    \frac{r_{<}}{r_{>}} \right)^n \\
  \times \int_0^{2 \pi} d \theta \int_0^{2 \pi} d \theta' \frac{- r R
    \sin(\theta) + r' R \sin(\theta') - r r' \sin(\theta - \theta')}{1 + s
    \cos(\theta)} \frac{r' + R \cos(\theta')}{1 + s' \cos(\theta')}
  \begin{pmatrix}
    \cos(\theta + \theta_R) \\ \sin(\theta + \theta_R)
  \end{pmatrix}
  \cos(n (\theta - \theta')),
\end{multline}
\end{widetext}
where $s = -2 r R/(r^2 + R^2)$ and $s'$ is defined correspondingly with $r$
replaced by $r'$. The next rather tedious steps, which can conveniently be
carried out in \textsc{Mathematica}, are to symmetrize the integrand with
respect to $\theta \to -\theta$ and $\theta' \to -\theta'$, and to rearrange the
trigonometric functions in the numerator such that the angular integrals can be
performed using the relation~\cite{Gradshteyn2007}
\begin{equation}
  \label{eq:tralala56}
  \int_0^{2 \pi} d \theta \frac{\cos(n \theta)}{1 + s \cos(\theta)} = \frac{2
    \pi}{\sqrt{1 - s^2}} \left( \frac{\sqrt{1 - s^2} - 1}{s} \right)^n,  
\end{equation}
which holds for $s^2 < 1$ and $n \geq 0$. In the resulting expression, the
summation over $n$ can be carried out, and finally, after performing the
integrals over $r$ and $r'$, magical cancellations lead to the result
$\mathbf{d}(\mathbf{R}) = 0$. Putting together Eqs.~\eqref{eq:tralala25}
and~\eqref{eq:tralala40} we finally obtain the second order correction to the
force in Eq.~\eqref{eq:5}.

\section{Superfluid density at the strong-coulping KPZ fixed point}
\label{sec:sf-density-kpz}

In this appendix, we explicitly evaluate the current-current response function
defined in Eq.~\eqref{eq:4} assuming that the fluctuations of the phase field
are governed by the strong coupling fixed point of the non-compact KPZ equation,
and hence ignoring the possible presence of topological defects. The superfluid
density is defined as the difference between the longitudinal and transverse
components of the current-current response function in the static limit, i.e.,
for vanishing frequency~\cite{Griffin1994,Hohenberg1965,Pitaevskii2003},
\begin{equation}
  \label{eq:sfd_4}
  \rho_s = \lim_{\mathbf{q} \to 0} \left( \chi_l(\mathbf{q}, 0) -
    \chi_t(\mathbf{q}, 0)\right).
\end{equation}
These components are defined in terms of the following decomposition of the
current-current response function, which is always possible in isotropic
systems:
\begin{equation}
  \label{eq:sfd_3}
  \chi_{ij}(\mathbf{q}, \omega) = \chi_l(\mathbf{q}, \omega) \frac{q_i q_j}{q^2}
  + \chi_t(\mathbf{q}, \omega) \left( \delta_{ij} -
    \frac{q_i q_j}{q^2} \right).
\end{equation}
Therefore, the superfluid density is just the coefficient of $q_i q_j/q^2$ in
the static current-current response function in the limit $\mathbf{q} \to 0$.

First we consider the contribution Eq.~\eqref{eq:sfd_45} to the current-current
response function, which after Fourier transformation becomes
\begin{equation}
  \label{eq:sfd_47}
  \chi_{ij}^{(1)}(Q) = D_0^2 q_i q_j G(Q).
\end{equation}
Here we denote the bare coefficient by $D_0$ in order to emphasize the
distinction from the renormalized one $D$. To keep the notation compact, we
denote $Q = \left( \mathbf{q}, \omega \right)$; $G(Q)$ is the retarded response
function,
\begin{equation}
  \label{eq:sfd_48}
  G(Q) \delta(Q + Q') = \langle \theta(Q) \hat{\theta}(Q') \rangle.
\end{equation}
Similarly, Fourier transformation of the second contribution to the
current-current response function given in Eq.~\eqref{eq:sfd_46}, which involves
the three-point function, yields
\begin{equation}
  \label{eq:sfd_49}
  \chi_{ij}^{(2)}(Q) = - D_0 \lambda q_i \int_{Q'} q_j' G_{112}(Q, Q'),
\end{equation}
where we set
$\int_Q = \int \frac{d^2 \mathbf{q}}{(2 \pi)^2} \frac{d \omega}{2 \pi}$.  Note
that we omit the subscript $\lambda_0$ for the KPZ non-linearity: as explained
in detail below, it is protected from renormalization by symmetries of the KPZ
equation. Our notation, which we choose for later convenience, indicates that
$G_{112}$ is the average value of a product of fields involving twice the phase
field $\theta_1 = \theta$ and once the response field
$\theta_2 = \hat{\theta}$.  Moreover, as in Eq.~\eqref{eq:sfd_48} we single
out a $\delta$-function that expresses invariance under spacial and temporal
translations and hence fixes the third argument in the Fourier transform of
$G_{112}$,
\begin{equation}
  \label{eq:sfd_50}
  G_{112}(Q_1, Q_2) \delta(Q_1 + Q_2 + Q_3) = \langle \theta(Q_1) \theta(Q_2) \hat{\theta}(Q_3) \rangle.
\end{equation}

As mentioned above, the superfluid density is determined by the contribution to
the current-current response function which is proportional to $q_i q_j/q^2$,
whereas the coefficient of $\delta_{ij}$ encodes the normal density. Thus, by
inspection of the momentum dependence in Eqs.~\eqref{eq:sfd_47}
and~\eqref{eq:sfd_49} we see that both $\chi^{(1)}$ and $\chi^{(2)}$ give
contributions to the superfluid density, and the normal density vanishes at the
present level of approximation. In fact, the present approach in which density
fluctuations are treated at lowest order, is analogous to keeping only the
zero-loop diagram in Fig.\ 1 of Ref.~\cite{Keeling2011} --- which still gives a
non-trivial result due to the non-equilibrium fluctuations of the phase at the
strong coupling fixed point of the KPZ equation. The leading contribution to the
normal density, however, is encoded in diagrams involving fluctuations of the
density at one-loop order.

As also mentioned in the main text, at the Gaussian fixed point corresponding to
a condensate in thermodynamic equilibrium, the contribution~\eqref{eq:sfd_49} to
the current-current response function evaluates to zero because all odd moments
of Gaussian distributed variables vanish; on the other hand, the retarded
response function in Eq.~\eqref{eq:sfd_47} reduces to its bare value
\begin{equation}
  \label{eq:sfd_51}
  G_0(Q) = \frac{i}{\omega + i D_0 q^2}.
\end{equation}
With Eq.~\eqref{eq:sfd_4}, we find the superfluid density in the Gaussian
approximation, $\rho_{s, 0} = D_0$ as expected. As pointed out in
Ref.~\cite{Keeling2011}, the crucial point leading to a finite value of the
superfluid density in the Gaussian approximation is the scaling of the bare
retarded response function with momentum as $G_0(\mathbf{q}, 0) \sim 1/q^2$,
which should be contrasted with the KPZ result
$G(\mathbf{q}, 0) \sim 1/q^{2 - \chi}$, obtained from the scaling analysis
described below around Eq.~\eqref{eq:sfd_53}, or from the explicit expression
Eq.~\eqref{eq:sfd_69} upon identifying the smallest possible momentum with the
inverse system size $q \sim 1/L$. Interestingly, if the expectation values in
Eqs.~\eqref{eq:sfd_47} and~\eqref{eq:sfd_49} are evaluated at the strong
coupling fixed point of the KPZ equation as we do in the following, it turns out
that the mechanism leading to a non-vanishing superfluid density in the
thermodynamic limit in Eq.~\eqref{eq:sfd_82} below is an entirely different one:
in fact, the contribution from the response function in Eq.~\eqref{eq:sfd_47}
vanishes for $L \to \infty$, while the three-point function in
Eq.~\eqref{eq:sfd_49} can be related to the non-linear term in the KPZ
equation~\eqref{eq:KPZ}, which is already hinted at by the observation that both
have exactly the same structure of derivatives and fields. The coupling
$\lambda$ of the non-linear vertex in the KPZ action Eq.~\eqref{eq:S_KPZ} is
protected from renormalization by symmetries of the KPZ
equation~\cite{Lebedev1994,Frey1994,Kamenev2011,Tauber2014a,Canet2010,Canet2011b,Canet2012}. Then,
the precise combination of $\lambda$ with powers of the renormalized values of
the diffusion constant $D$ and the noise strength $\Delta$ that appear in the
evaluation of Eq.~\eqref{eq:sfd_49} gives just the dimensionless KPZ coupling
$g = \lambda^2 \Delta/D^3$, which takes a universal value $g_{*}$ at the strong
coupling fixed point~\cite{Canet2010,Canet2011b,Canet2012,Kloss2012}, leading to
the contribution to the superfluid density~\eqref{eq:sfd_82}, which remains
finite even in the thermodynamic limit. In other words, whereas in the Gaussian
approximation the contribution to the superfluid density due to
Eq.~\eqref{eq:sfd_47} is finite while the one from the three-point function in
Eq.~\eqref{eq:sfd_49} vanishes for $L \to \infty$, non-equilibrium fluctuations
at the strong coupling fixed point of the KPZ equation lead to exactly the
opposite conclusion.

\subsection{Scaling analysis}
\label{sec:scaling-analysis}

Before going into the details of the calculation of the superfluid density, let
us show that this conclusion can already be drawn from a simple scaling analysis
for the two contributions $\chi^{(1)}$ and $\chi^{(2)}$ to the current-current
response function. We count momentum dimensions, i.e., $[q] = 1$ and for the
integration measures of time and space we find $[d t] = - z$ with the dynamical
exponent $z$ and $[d^d \mathbf{r}] = - d$ in $d$ spatial dimensions. The scaling
dimension of the phase field is the roughness exponent, $[\theta(X)] = - \chi$,
and we denote the scaling dimension of the response field as
$[\hat{\theta}(X)] = - \hat{\chi}$. Then, the Fourier transform of the
contribution to the current-current response function in Eq.~\eqref{eq:sfd_45}
scales as
\begin{equation}
  \label{eq:sfd_53}
  [\chi_{ij}^{(1)}(\mathbf{q}, 0)] = - z - d + 2 - \chi - \hat{\chi} = \chi,
\end{equation}
where we used that $[\partial/\partial x] = [q] = 1$, and the second equality
follows from the scaling relations $d + \chi + \hat{\chi} = 0$ and
$\chi + z = 2$~\cite{Kamenev2011} which in turn follow from symmetries of the
KPZ equation. Therefore, $\chi^{(1)}$ yields a contribution to the superfluid
density that scales as $\rho_s^{(1)} \sim L^{-\chi}$ (note that
$[L] = [1/q] = -1$ so that Eq.~\eqref{eq:sfd_53} indeed implies
$\chi_{ij}^{(1)}(\mathbf{q}, 0) \sim L^{-\chi}$).  In the same way, we can see
that the second contribution to the current-current response function, given in
Eq.~\eqref{eq:sfd_46}, has a vanishing scaling dimension,
\begin{equation}
  \label{eq:sfd_54}
  [\chi_{ij}^{(2)}(\mathbf{q}, 0)] = -  z - d + 2 - 2 \chi - \hat{\chi} = 0.
\end{equation}
Here we used the same scaling relations as above. These considerations show that
a driven-dissipative condensate that is described by the KPZ equation indeed
should be expected to have a finite superfulid density if vortices are not taken
into account. However, to obtain an explicit expression for the superfulid
density, we have to evaluate the correlation functions\footnote{Response
  functions are correlation functions involving response fields.} in
Eqs.~\eqref{eq:sfd_47} and~\eqref{eq:sfd_49}.

\subsection{Evaluation of $\rho_s$ at the strong coupling fixed point}
\label{sec:evaluation-rho_s-scfp}

Our strategy is to first express these response functions in terms of
irreducible vertex functions~\cite{Negele1998} and then approximate the latter
by their low frequency and momentum expansions, the form of which is strongly
restricted by the Ward identities associated with the various symmetries of the
KPZ equation.

Here and in the following we denote the two-point response and
correlation functions by $G$ and $C$, respectively, and we denote $X = \left(
  \mathbf{r}, t \right)$,
\begin{equation}
  \label{eq:sfd_64}
  \begin{split}
    G(X - X') & = G_{12}(X, X') = \langle \theta_1(X) \theta_2(X') \rangle \\ &
    = \langle \theta(X) \hat{\theta}(X') \rangle, \\ C(X - X') & = G_{11}(X,
    X') = \langle \theta_1(X) \theta_1(X') \rangle \\ & = \langle \theta(X) \theta(X') \rangle.
  \end{split}
\end{equation}
Note that $G_{22}$ vanishes due to causality~\cite{Kamenev2011,Altland2010a}. In
terms of vertex functions, i.e., derivatives of the effective action
$\Gamma$~\cite{Negele1998}, the response and correlation functions can be
expressed as
\begin{equation}
  \label{eq:sfd_65}
  \begin{split}
    G(Q) & = 1/ \Gamma_{12}(-Q), \\
    C(Q) & = - \Gamma_{22}(Q) /( \Gamma_{12}(Q) \Gamma_{12}(-Q) ).
  \end{split}
\end{equation}
For the vertex functions we are using the notation
\begin{multline}
  \label{eq:16}
  \Gamma_{i_1, i_2, i_3, \dotsc}(X_1, X_2, X_3, \dotsc) \\ = \frac{\delta^n
    \Gamma}{\delta \varphi_{i_1}(X_1) \delta \varphi_{i_2}(X_2) \delta
    \varphi_{i_3}(X_3) \dotsb},
\end{multline}
where $\varphi_i(\mathbf{r}, t) = \langle \theta_i(\mathbf{r}, t)
\rangle$. After Fourier transformation, we single out a frequency and momentum
conserving $\delta$-function as described below Eq.~\eqref{eq:sfd_49}.

From Eq.~\eqref{eq:sfd_65} we obtain approximate expressions for the response
and correlation functions by inserting for the vertex functions the respective
low-frequency and low-momentum expansions. In the case of $\Gamma_{12}(Q)$, the
form of this expansion is restricted by the Ward identity associated with the
shift-gauged symmetry of the KPZ
equation~\cite{Lebedev1994,Canet2011b,Canet2012}: indeed, this symmetry entails
that the coefficient of the term $\int_X \hat{\theta} \partial_t \theta$ in the
KPZ action~\eqref{eq:S_KPZ} is not renormalized~\cite{Frey1994}. Hence, we have
for arbitrary frequencies and at zero momentum the exact relation
\begin{equation}
  \label{eq:sfd_66}
  \Gamma_{12}(0, \omega) = i \omega.
\end{equation}
For finite momentum, rotational invariance implies that the lowest order
contribution to an expansion in powers of $\mathbf{q}$ is proportional to
$q^2$. This leads to
\begin{equation}
  \label{eq:sfd_67}
  \Gamma_{12}(Q) = i \omega + D q^2 + O(\omega q^2, q^4).
\end{equation}
At the strong-coupling fixed point, the coefficient $D$ obeys the finite-size
scaling $D \sim D_{*} L^{\chi}$~\cite{Kamenev2011}, where $D_{*}$ is a
non-universal constant. For the $\Gamma_{22}$ vertex there is no restriction
from the shift-gauged symmetry and, therefore, its leading contribution in the
limit of vanishing frequency and momentum is just a constant,
\begin{equation}
  \label{eq:sfd_68}
  \Gamma_{22}(Q) = - 2 \Delta + O(\omega, q^2),
\end{equation}
which scales with system size as
$\Delta \sim \Delta_{*} L^{3 \chi + d - 2}$~\cite{Kamenev2011}. Plugging
Eqs.~\eqref{eq:sfd_67} and~\eqref{eq:sfd_68} into Eq.~\eqref{eq:sfd_65} yields the
low-frequency and low-momentum scaling forms of the response and correlation
functions
\begin{equation}
  \label{eq:sfd_69}
  \begin{split}
    G(Q) & = \frac{i}{\omega + i D q^2}, \\ C(Q) & = 2 \Delta \abs{G(Q)}^2 =
    \frac{2 \Delta}{\omega^2 + D^2 q^4}.
  \end{split}
\end{equation}

For the three-point function appearing in the current-current response function
Eq.~\eqref{eq:sfd_49}, the relation corresponding to Eq.~\eqref{eq:sfd_65} reads
\begin{multline}
  \label{eq:sfd_71}
  G_{112}(Q, Q') = - \left( \Gamma_{122}(-Q - Q', Q) G(Q) G(Q') \right. \\ +
  \Gamma_{112}(- Q - Q', Q) C(Q) G(Q') \\ \left. + \Gamma_{112}(Q', - Q - Q')
    C(Q') G(Q) \right) G(Q + Q').
\end{multline}
In order to make progress with Eq.~\eqref{eq:sfd_71} we have to specify the
vertex functions. As above, we restrict ourselves to the lowest order in
frequency and momentum and consider the following ansatz,
\begin{equation}
  \label{eq:sfd_72}
  \Gamma_{112}(Q, Q') = \gamma_1 + \gamma_2 \left( \omega + \omega' \right) + \gamma_3 \mathbf{q} \cdot \mathbf{q}' +
  \gamma_4 \left( q^2 + q^{\prime 2} \right),
\end{equation}
which incorporates rotational invariance and symmetry of $\Gamma_{112}(Q, Q')$
under exchange of its arguments, as follows from the commutativity of the
functional derivatives with respect to $\varphi(Q)$ and $\varphi(Q')$. The
shift-gauged symmetry of the KPZ action
implies~\cite{Lebedev1994,Canet2011b,Canet2012}
\begin{equation}
  \label{eq:sfd_73}
  \Gamma_{112}(0, \omega, \mathbf{q}', \omega') = \gamma_1 + \gamma_2
  \left( \omega + \omega' \right) + \gamma_4 q^{\prime 2} = 0,
\end{equation}
leading to $\gamma_1 = \gamma_2 = \gamma_4 = 0.$ We are left with a single
parameter $\gamma_3$, which is in fact determined by another symmetry of the KPZ
action: the ansatz~\eqref{eq:sfd_72} leads to the first equality in
\begin{equation}
  \label{eq:sfd_74}
  \begin{split}
    \gamma_3 & = \frac{1}{d} \left. \nabla_{\mathbf{q}} \cdot
      \nabla_{\mathbf{q}'} \Gamma_{112}(Q, Q') \right\rvert_{Q = Q' = 0} \\ & =
    - i \lambda \frac{\partial}{\partial \omega} \left. \Gamma_{12}(0, \omega)
    \right\rvert_{\omega = 0} = \lambda,
  \end{split}
\end{equation}
whereas in the second one we used the Ward identity associated with the Galilean
symmetry of the KPZ equation~\cite{Lebedev1994,Canet2011b,Canet2012} in order to
express the derivatives with respect to momenta of the vertex $\Gamma_{112}$ in
terms of a derivative with respect to frequency of the lower order vertex
function $\Gamma_{12}$, for which we then inserted Eq.~\eqref{eq:sfd_66}. Thus
we have
\begin{equation}
  \label{eq:sfd_75}
  \Gamma_{112}(\mathbf{q}, \omega, \mathbf{q}', \omega') = \lambda \mathbf{q}
  \cdot \mathbf{q}',
\end{equation}
which is again just the bare vertex already present in the KPZ
action~\eqref{eq:S_KPZ}. Renormalization of this vertex might occur only at
higher orders in an expansion in powers of frequency and momentum.

In analogy to Eq.~\eqref{eq:sfd_72}, for the vertex $\Gamma_{122}(Q, Q')$ we
start with the following ansatz, taking into account that it has to be symmetric
under the exchange of $Q'$ by $-Q-Q'$:
\begin{equation}
  \label{eq:17}
  \Gamma_{122}(Q, Q') = \kappa_0 + \kappa_1 \omega + \kappa_2 q^2 + \kappa_3
  \left( \mathbf{q} + \mathbf{q}' \right) \cdot \mathbf{q}'.
\end{equation}
Note that a term linear in $\omega'$ is forbidden by the above-mentioned
symmetry. From the shift-gauged symmetry it follows that
$\kappa_0 = \kappa_1 = \kappa_3 = 0$. However, the presence of a term
$\propto q^2$ with an unknown coefficient $\kappa_2$ cannot be excluded.

With the expressions for the response, correlation, and vertex functions, we
proceed to evaluate the two contributions to the current-current response
function, Eqs.~\eqref{eq:sfd_47} and~\eqref{eq:sfd_49}, at vanishing frequency. We find
\begin{equation}
  \label{eq:sfd_77}
  \chi_{ij}^{(1)}(\mathbf{q}, 0) = \frac{D_0^2}{D} \frac{q_i q_j}{q^2},
\end{equation}
leading to a contribution to the superfluid density
\begin{equation}
  \label{eq:sfd_78}
  \rho_s^{(1)} = \frac{D_0^2}{D} \sim \frac{D_0^2}{D_{*}} L^{-\chi},
\end{equation}
which vanishes in the thermodynamic limit $L \to \infty$. Note that precisely
this contribution involving the two-point response function remains finite in a
Gaussian approximation in which $G_0(\mathbf{q}, 0) \sim 1/q^2$. Let us turn now
to the evaluation of Eq.~\eqref{eq:sfd_49}, which can be rewritten as
\begin{multline}
  \label{eq:sfd_79}  
  \chi_{ij}^{(2)}(\mathbf{q}, 0) = D_0 \lambda q_i \int_{Q'} q_j' \left[ \kappa_2
    \abs{\mathbf{q} - \mathbf{q}'}^2 G(\mathbf{q}, 0) G(\mathbf{q}', \omega')
  \right. \\ - \lambda \left( \mathbf{q} + \mathbf{q}' \right) \cdot \left(
    \mathbf{q} C(\mathbf{q}, 0) G(\mathbf{q}', \omega') \right. \\
  \left. \left. + \mathbf{q}' G(\mathbf{q}, 0) C(\mathbf{q}', \omega') \right)
    \vphantom{\abs{\mathbf{q} - \mathbf{q}'}^2} \right] G(\mathbf{q} +
  \mathbf{q}', \omega').
\end{multline}
The integral over $\omega'$ of the product
$G(\mathbf{q}', \omega') G(\mathbf{q} + \mathbf{q}', \omega')$ vanishes since
the integrand has poles only in the lower half-plane; hence there is no
contribution $\propto \kappa_2$ and also the first term $\propto \lambda$ does not
contribute to the current-current response function. For the remaining integral
over frequency we find
\begin{equation}
  \label{eq:sfd_80}
  \int_{\omega'} C(\mathbf{q}', \omega') G(\mathbf{q} + \mathbf{q}', \omega') =
  \frac{\Delta}{D^2 q^{\prime 2}} \frac{1}{q^{\prime 2} + \abs{\mathbf{q} + \mathbf{q}'}^2}.
\end{equation}
The integral over the momentum $\mathbf{q}'$ can be carried out exactly in 2D
and we obtain
\begin{equation}
  \label{eq:sfd_81}
  \chi_{ij}^{(2)}(\mathbf{q}, 0) = - \frac{\ln 2}{8 \pi} \frac{D_0 \lambda^2
    \Delta}{D^3} \frac{q_i q_j}{q^2},
\end{equation}
which yields the contribution to the superfluid density
\begin{equation}
  \label{eq:sfd_82}
  \rho_s^{(2)} = - \frac{\ln 2}{8 \pi} \frac{D_0 \lambda^2
    \Delta}{D^3} \sim \frac{\ln 2}{8 \pi} D_0 g_{*},
\end{equation}
where $g_{*}$ is the value of the dimensionless KPZ coupling
$g = \lambda^2 \Delta/D^3$ at the strong coupling fixed
point~\cite{Kamenev2011,Canet2010,Canet2011b,Canet2012}.  Quantitatively, this
result must be considered as a rough estimate, as it is based on the low
frequency and momentum form of the correlation function \eqref{eq:sfd_69}, and
there can be corrections resulting from the deviation of these forms at higher
frequencies and momenta. However, the calculation demonstrates explicitly the
finiteness of the superfluid stiffness: possible quantitative corrections would
be non-universal (not dependent on the fixed point coupling) and thus are not
expected to cancel the finite contribution that we identified. The sum of
Eqs.~\eqref{eq:sfd_78} and~\eqref{eq:sfd_82} yields the final result for the
superfluid density in the absence of topological defects within our
approximation:
\begin{equation}
  \label{eq:11}
  \rho_s = \rho_s^{(1)} + \rho_s^{(2)} = \frac{D_0^2}{D_{*}} L^{-\chi} +
  \frac{\ln 2}{8 \pi} D_0 g_{*}.
\end{equation}

\bibliography{KPZduality}

\end{document}